\documentclass[twocolumn]{aastex701}
\usepackage{color}
\usepackage[titletoc]{appendix}
\usepackage{amsmath}
\usepackage{amssymb}
\usepackage{mathtools}
\usepackage{upgreek}
\usepackage{float}
\usepackage{comment}
\usepackage{enumitem}
\usepackage{natbib}
\usepackage{graphicx}
\usepackage{bm}
\usepackage{totcount}
\usepackage{multirow}
\usepackage{pifont}
\usepackage{makecell}
\usepackage{hyperref}

\renewcommand\cellgape

\newtotcounter{citnum} 
\def\oldbibitem{} \let\oldbibitem=\bibitem
\def\bibitem{\stepcounter{citnum}\oldbibitem}

\shortauthors{Sethi et. al.}
\shorttitle{Hydrodynamic Atmospheric Escape in Planets}
\begin{document} 
\title{Atmospheric Escape Rates of Planets in Stellar Tidal Fields from 3-D Hydrodynamic Simulations}


\author[0000-0002-6576-3346]{Ritika Sethi}
\affiliation{Department of Physics, Massachusetts Institute of Technology, Cambridge, MA 02139, USA}
\affiliation{MIT Kavli Institute for Astrophysics and Space Research, Massachusetts Institute of Technology, Cambridge, MA 02139, USA}
\email{rsethi@mit.edu}
\author[0000-0002-1417-8024]{Morgan MacLeod}
\affiliation{Institute for Theory \& Computation, Center for Astrophysics, Harvard \& Smithsonian, Cambridge, MA 02138, USA}
\email{morgan.macleod@cfa.harvard.edu}
\author[0000-0003-3130-2282]{Sarah Millholland}
\affiliation{Department of Physics, Massachusetts Institute of Technology, Cambridge, MA 02139, USA}
\affiliation{MIT Kavli Institute for Astrophysics and Space Research, Massachusetts Institute of Technology, Cambridge, MA 02139, USA}
\email{email}

\begin{abstract}
Thermally driven atmospheric escape, including photo-evaporation and core-powered mass-loss, plays a key role in shaping the evolution of close-in exoplanets, yet most current models rely on simplified one-dimensional descriptions of atmospheric escape. In this work, we perform 3D hydrodynamic simulations of atmospheric outflows from a Jupiter-sized planet embedded in the gravitational potential of a solar-type host star, and compare these results with 1D models to identify the regimes where they perform well and where they break down. We explore a range of configurations by varying the degree of Roche-lobe filling and the thermal state of the outflow. We find that systems with weak tidal influence and high-temperature winds produce nearly spherical and isotropic outflows, whereas more Roche-lobe-filling and cooler winds develop strong anisotropy and form two-tailed structures. We show that the commonly used 1D Parker wind model performs well only in the weak-tides regime, while including tidal corrections yields reasonable estimates of mass-loss rates and captures the mean radial density profile across all regimes, but fails to reproduce the intrinsically three-dimensional, angle-dependent nature of the flow as the outflow transitions from spherical to tidally structured tails. Motivated by these results, we develop a physically informed \textit{Mixture Model}, calibrated using our 3D simulations, that accurately predicts mass-loss rates across the parameter space explored and outperforms 1D model with tidal corrections.

\end{abstract}

\section{Introduction} \label{sec: Introduction} 
Hydrodynamic escape plays a central role in shaping the atmospheric evolution of close-in exoplanets, influencing their physical structure and composition. Thermal mass loss driven by stellar irradiation, i.e., \textit{photoevaporation} \citep{2003ApJ...598L.121L} or by internal cooling of the planetary core, known as \textit{core-powered mass loss} \citep{2018MNRAS.476..759G, 2019MNRAS.487...24G}, has been widely invoked to explain key demographic features of short-period planet populations. These include the radius valley separating super-Earths from sub-Neptunes near $\sim 1.5-2 R_\oplus$ \citep{2013ApJ...775..105O, 2013ApJ...776....2L,2017AJ....154..109F, 2017ApJ...847...29O} and the hot Neptune desert, characterized by a dearth of Neptune-sized planets at short orbital periods \citep{2016A&A...589A..75M, 2025AJ....169..117V}. Moreover, there is evidence that strongly irradiated gas giants experience Roche lobe outflow that removes substantial atmospheric mass over $\sim$Gyr timescales \citep{2015ApJ...813..101V, 2019A&A...624A.101L, 2019AREPS..47...67O, 2023AJ....165..200S, 2026ApJ...997..139H}.  

Alongside demographic features, atmospheric outflows can now be directly detected in close-in transiting planets through transit spectroscopy, via absorption in strong atomic lines such as $\rm Ly\alpha$, $\rm H\alpha$, and the metastable helium line (He*) at 1083 nm \citep{2003Natur.422..143V, 2018Natur.557...68S, 2018ApJ...855L..11O, 2019AREPS..47...67O, 2023IAUS..370...56D, 2024RvMG...90..411K, 2026AJ....171..257S}. These observations provide an important opportunity to test and refine current mass-loss models.

Reliable estimates of mass-loss rates are therefore essential for modeling the long-term structural and dynamical evolution of close-in planetary systems \citep{2012MNRAS.425.2931O, 2013ApJ...766L..20F, 2021MNRAS.508.5886R}. However, the mass-loss rates used in population-level studies remain uncertain, often by orders of magnitude. A key reason is that many studies adopt highly simplified descriptions of atmospheric escape. In the commonly used energy-limited framework, the mass-loss rate is tied to the incident high-energy stellar flux and an assumed heating efficiency, effectively balancing the input stellar energy against the gravitational binding energy of the atmosphere \citep{1981Icar...48..150W, 2016MNRAS.460.1300E}. While, this provides a useful scaling for the mass-loss rate, it does not capture the dynamical structure of the outflow. 

More physically motivated models are based on Parker-type wind solutions, in which gas pressure accelerates the flow from subsonic to supersonic velocities through a critical sonic point \citep{1960ApJ...132..821P}. These models treat atmospheric escape as a one-dimensional (1D) hydrodynamic outflow, solving for the acceleration profile, sonic point location, and radial density and velocity structure. Building on this framework, several studies have developed highly sophisticated 1D models of hydrodynamic escape. Early models incorporated detailed photochemistry, heating/cooling, and ionization balance for hot-Jupiter atmospheres \citep{2004Icar..170..167Y, 2007P&SS...55.1426G, 2009ApJ...693...23M}. More recent work has extended these models to include multi-species escape, atomic metal chemistry, radiative transfer, and more detailed photoionization treatments \citep{2018A&A...619A.151K, 2021RNAAS...5...74K, 2022ApJ...929...52K, 2023ApJ...951..123H, 2023A&A...675A.193L, 2024A&A...684A..26K}. These physical processes are implemented in several publicly available codes such as ATES \citep{2021A&A...655A..30C, 2022A&A...663A.122C}, AIOLOS \citep{2023MNRAS.523..286S}, \texttt{p-winds} \citep{2022A&A...659A..62D, 2022ApJ...927...96V}, \texttt{sunbather} \citep{2024A&A...688A..43L}, and Wind-AE \citep{2025ApJ...995..198B}. 

Some of the most studied planets orbit close to their host star, so tidal gravity is important. In 1D models, tidal effects are included by replacing atmospheric escape to infinity with escape to the Roche-lobe boundary, thereby lowering the effective potential barrier \citep{2007A&A...472..329E} but the underlying assumption of spherical symmetry remains. However, 3D simulations have shown that this assumption may break down as a planet approaches Roche-lobe filling, where gas can preferentially escape through low-barrier channels that open near the $L_1$ and $L_2$ Lagrange points \citep{2013A&A...557A.124B, 2019MNRAS.483.1481D, 2023SciA....9F8736Z, 2024A&A...692A.230C, 2025NatCo..1610822A, 2025ApJ...988...63M}. Under such conditions, the Lagrange points effectively act as nozzle-like throats, producing anisotropic, tidally directed outflows and extended tails. Some studies \citep[e.g.,][]{2015ApJ...808..173T, 2016ApJ...820....3C, 2017MNRAS.466.2458C, 2024AJ....167..142G, 2024A&A...684A..20N, 2025A&A...695A.186N, 2025ApJ...988...63M}, also show that escaping atmospheres can develop pronounced day–night and leading–trailing asymmetries. 


More advanced and fully self-consistent 3D models have also been developed in recent years that include photoheating, radiative cooling, and detailed thermochemistry \citep[e.g.,][]{2015A&A...576A..21S, 2016A&A...586A..75S, 2018ApJ...860..175W}. While these models provide a more complete description of atmospheric escape, they are computationally expensive and are not feasible for large scale population-level studies or coupling with long-term orbital evolution. This motivates the need to assess the validity of simpler 1D models, specifically to determine the regimes in which they remain applicable and the physical processes they fail to capture. 

In this work, we address this gap by providing a systematic suite of 3D hydrodynamic simulations, isolating tidal effects from thermodynamic processes to quantify their impact on both mass-loss rates and flow morphology. We simulate a Jupiter-sized planet embedded in the gravitational potential of a solar-mass star, exploring a broad range of thermal states and degrees of Roche-lobe filling (45 models in total). The thermal state of the flow is parametrized through the sound speed, allowing us to remain agnostic to the underlying heating mechanism. This wide parameter space allows us to map how escape morphology varies with atmospheric energy budget and tidal strength. Our simulations also provide a controlled benchmark for commonly used 1D atmospheric escape models. In particular, we identify the regimes in which these models remain applicable despite their assumption of spherical symmetry, and the regimes in which stronger tidal effects lead to significant deviations in both the flow geometry and the resulting mass-loss rates.

This paper is organized as follows. In \S \ref{sec:methods}, we describe the numerical setup of our 3D hydrodynamic simulations, including the governing equations, boundary conditions, and computational domain. \S \ref{sec-results} presents the results, exploring the morphology of the outflow across the parameter space, including the transition from subsonic to supersonic flow, the structure of the sonic surface, the breakdown of spherical symmetry. We also analyze the angular mass-flux distribution. In \S \ref{sec:mass_loss rate models} we discuss several 1D models for mass-loss rate prediction, and introduce the \textit{Mixture Model}. In \S \ref{sec:discussion}, we examine the regimes in which 1D models perform well and where they fail, compare the \textit{Mixture Model} to baseline 1D predictions across the observed exoplanet population, and discuss differences in radial density profiles between 1D models and those obtained from 3D simulations. Finally, we summarize our findings in \S \ref{sec:conclusions}.

\section{3-D Hydrodynamic Simulations}  \label{sec:methods}
The physical configuration of our simulation setup corresponds to a corotating star–planet system, though the computational domain is limited to the planetary vicinity and does not explicitly model the star. We model the dynamics of gas outflowing from the surface of the planet, representing its escaping atmosphere. No stellar wind or mass outflow from the star is included in the simulation. All simulations are carried out using the Athena++ code \citep{2020ApJS..249....4S}, a modern Eulerian (magneto) hydrodynamics framework that builds upon the original Athena code \citep{2008ApJS..178..137S}.

Our models are performed in units where $G = M = a = 1$. Here, $G$ is the gravitational constant, $M = M_\star + M_{\rm p} = 1$ is the total mass of the system, with the stellar and planetary components having masses, $M_\star = 0.999$, and $M_{\rm p} = 0.001$ in code units, respectively; and $a$ is the orbital separation of the system's circular orbit. Under this normalization, the unit velocity is $\sqrt{GM/a} = 1$, and the unit time is $\sqrt{a^3/GM} = 1$, such that the orbital period ($P_{\rm orb}$) is $2\pi$. This setup corresponds to a system consisting of roughly a Jupiter-mass planet orbiting a solar-mass star.

\subsection{Setup Overview}
While magnetic stresses are certainly important in some regions of planetary outflows and interactions \citep{2009ApJ...704L..85C,2020MNRAS.494.5044T, 2020ApJ...890...88O, 2024MNRAS.527.5117S}, we adopt a significant simplification in our modeling by neglecting magnetohydrodynamic effects and focus solely on the purely hydrodynamic components of planetary outflows. 

Our methodology is a modified version of that described by \citet{2020ApJ...902...85M}, who modeled hydrodynamic winds in twin-star binaries. We have adapted their code to represent a planet–star system by disabling stellar wind injection, adjusting the mass ratio, and modifying the planetary surface boundary conditions to account for the Roche potential, resulting in a tidally distorted geometry for the planet that follows contours of effective gravitational potential. The planet and star are placed along the $x$-axis in the rotating frame, with the planet located at $(x_p, y_p, z_p) = (M_\star/M, 0, 0)$, and star at $(x_\star, y_\star, z_\star) = (-M_p/M, 0, 0)$, such that the system's center of mass lies at the origin.

\subsubsection{Equations Solved}
Our hydrodynamic models solve the equations of inviscid gas dynamics in a rotating reference frame centered on the system’s center of mass. The rotation frequency of this frame is equal to the orbital frequency of the planet, $\Omega = \sqrt{GM/a^3}$. The simulations are carried out on a Cartesian mesh that includes one level of static mesh refinement surrounding the planet. The conservation equations solved are:
\begin{subequations}\label{eq:hydrodynamics}
    \begin{align}
        \partial_t\rho + \nabla \cdot (\rho \boldsymbol{\nu}) &= 0, \\  
        \partial_t(\rho\boldsymbol{\nu}) + \nabla \cdot (\rho \boldsymbol{\nu \nu}
        + P\boldsymbol{I}) &= -\rho\boldsymbol{a}_{\rm ext}, \\
        \partial_tE + \nabla \cdot [(E + P)\boldsymbol{\nu}] &= -\rho\boldsymbol{a}_{\rm ext}\cdot \boldsymbol{\nu}        
    \end{align}
\end{subequations}%
expressing the mass continuity, momentum evolution, and energy evolution, respectively. In the above equations, $\rho$ is the mass density, $\rho\boldsymbol{\nu}$ is the momentum density, and $E = \epsilon + \rho \boldsymbol{\nu}\cdot\boldsymbol{\nu}/2$ is the total energy density, where $\epsilon$ is the internal energy density. $P$ is the pressure, $\boldsymbol{I}$ is the identity tensor, and $\boldsymbol{a}_{\rm ext}$ is the external acceleration associated with the rotating frame of reference. These equations are closed by an ideal gas equation of state, $P = (\gamma - 1)\epsilon$, where $\gamma = 1.01$ is the adiabatic index of the gas. This simplified equation of state is not intended to realistically capture the thermodynamics of planetary outflows; rather, it ensures that along adiabats the gas nearly remains isothermal. The source terms of the star and planet system gravity  and rotating reference frame are contained in the acceleration,
 
\begin{equation} \label{eq:source term}
    \boldsymbol{a}_{\rm ext} = -\frac{GM_\star}{\left| \boldsymbol{r}_\star \right|^3}
    \boldsymbol{r}_\star - \frac{GM_p}{\left| \boldsymbol{r}_p \right|^3}\boldsymbol{r}_p - \boldsymbol{\Omega} \times \boldsymbol{\Omega} \times \boldsymbol{r} - 2\boldsymbol{\Omega} \times \boldsymbol{\nu},
\end{equation}
where $\boldsymbol{r}$ is the position vector from origin to a given zone, and $\boldsymbol{r}_p$, $\boldsymbol{r}_\star$ are the vectors from the planetary and stellar centers to a zone, respectively. The angular momentum of the orbit is aligned with the $+z$ direction, so the orbital frequency vector is $\boldsymbol{\Omega} = (0, 0, \Omega)$ and the $(x,y)$ plane therefore defines the planet’s orbital plane.

\subsubsection{Hydrodynamic Escape Parameter}
For an isolated planet, thermal escape is governed by the competition between the thermal energy of gas in the upper atmosphere and the planet's gravitational binding energy. Thermal escape has two limiting regimes: Jeans escape, in which particles escape molecule by molecule from a collisionless exosphere, and hydrodynamic escape, in which sufficiently strong heating and frequent collisions allow the atmosphere to respond as a bulk outflow \citep{2020JGRA..12527639G}. The relevant thermal state is therefore that of the upper atmosphere or exosphere, whose temperature depends not only on orbital distance, but also on the incident high-energy stellar flux, atmospheric composition, and associated heating/cooling processes \citep{2004Icar..170..167Y}. 
In this work, we do not model these microphysical heating, cooling, or chemical processes explicitly. 
Instead, we parametrize the thermal state of escaping gas through the sound speed, $c_s$. 

The binding energy per unit mass is $GM_p/R_p$, where $R_p$ is the planetary radius, while the thermal energy per unit mass is on the order of $\sim c_s^2$, where $c_s$ is the gas sound speed. The ratio between these two energy scales defines the hydrodynamic escape parameter ($\lambda_p$) given by,
\begin{equation} \label{eq:lambdap}
    \lambda_p = \frac{GM_p}{c_s^2R_p}.
\end{equation}
A small $\lambda_p$ indicates that the thermal energy dominates over the gravitational binding energy, allowing a hydrodynamic wind with a characteristic velocity, $\nu_w \sim c_s$ to emerge. Conversely, a value of $\lambda_p \gg 1$ corresponds to a deep gravitational potential well in which the gas is strongly confined \citep{1960ApJ...132..821P, 1999isw..book.....L, 2007A&A...461.1185L}. In the limit of sufficiently large $\lambda_p$, thermal escape is expected to transition from hydrodynamic outflow to Jeans escape \citep{2020JGRA..12527639G}. Here, we focus only on the hydrodynamic regime and assume that the gas remains collisional, such that the gas mean free path, $l \ll R_p$.


Additionally, planetary outflows in real systems exist within the combined star–planet gravitational potential \citep{2007A&A...472..329E, 2017MNRAS.466.2458C, 2019ApJ...873...89M, 2023MNRAS.518.4357O, 2025ApJ...988...63M}. It is therefore important to include the gravitational influence of both the bodies, as well as the rotational terms (see Eq. \ref{eq:source term}). The planet’s gravity dominates within its Hill sphere (or Roche lobe), which has an approximate radius given by
\begin{equation}
    r_{\rm H} = \left(\frac{M_p}{3M_\star}\right)^{1/3}a,
\end{equation}
which defines the region where the planet’s gravitational potential is dominant, with a corresponding potential depth, $\Phi_{\rm Hill}$ of $-GM_p/r_{\rm H}$. Analogously to $\lambda_p$, we define the hydrodynamic escape parameter of the Hill sphere as
\begin{equation} \label{eq:lambda_hill}
    \lambda_{\rm Hill} = \frac{GM_p}{c_s^2 r_{\rm H}}.
\end{equation}
This formulation links local thermal conditions, through $c_s$, to the large-scale escape geometry determined by the star–planet potential (see also \S 2 of \citet{2024arXiv241112895M} for a similar treatment). The choice of $c_s$ therefore controls the planet's atmospheric pressure and the efficiency of hydrodynamic escape in our models.

\begin{table}  
    \centering
    \setlength{\tabcolsep}{3.5pt}
    \begin{longtable}{>{\footnotesize}l >{\footnotesize}l >{\footnotesize}l >{\footnotesize}l >{\footnotesize}l >{\footnotesize}l >{\footnotesize}l}
    \hline
    \hline
        $f_\phi$ & $\lambda_{\rm Hill}$ & $\lambda_p$ & $\dot M_{\rm sim}$ & $\dot M_{\rm PW}$ & $\dot M_{\rm nozzle}$ & $ \dot M_{\rm mix}$  \\ \hline 
        0.010 & 1.606 & 2.50 & 1.00 $\times 10^{-4}$ & 1.32 $\times 10^{-3}$ & 1.78 $\times 10^{-3}$ & 9.86 $\times 10^{-4}$ \\ 
        0.016 & 1.600 & 2.50 & 9.92 $\times 10^{-4}$ & 1.31 $\times 10^{-3}$ & 1.77 $\times 10^{-3}$ & 9.78 $\times 10^{-4}$ \\ 
        0.024 & 1.590 & 2.50 & 9.80  $\times 10^{-4}$ & 1.30 $\times 10^{-3}$ & 1.76 $\times 10^{-3}$ & 9.67 $\times 10^{-4}$ \\ 
        0.038 & 1.573 & 2.50 & 9.64 $\times 10^{-4}$ & 1.28 $\times 10^{-3}$ & 1.76 $\times 10^{-3}$ & 9.49 $\times 10^{-4}$ \\ 
        0.059 & 1.551 & 2.50 & 9.35 $\times 10^{-4}$  & 1.25 $\times 10^{-3}$ & 1.74 $\times 10^{-3}$ & 9.23 $\times 10^{-4}$ \\ 
        0.092 & 1.515 & 2.50 & 8.99 $\times 10^{-4}$ & 1.21 $\times 10^{-3}$ & 1.72 $\times 10^{-3}$ & 8.85 $\times 10^{-4}$ \\ 
        0.143 & 1.465 & 2.50 & 7.91 $\times 10^{-4}$ & 1.15 $\times 10^{-3}$ & 1.68 $\times 10^{-3}$ & 8.34 $\times 10^{-4}$ \\ 
        0.224 & 1.393 & 2.50 & 7.81 $\times 10^{-4}$ & 1.07 $\times 10^{-3}$ & 1.65 $\times 10^{-3}$ & 7.66 $\times 10^{-4}$ \\ 
        0.349 & 1.297 & 2.50 & 6.94 $\times 10^{-4}$ & 9.67 $\times 10^{-4}$ & 1.59 $\times 10^{-3}$ & 6.82 $\times 10^{-4}$ \\ 
        0.543 & 1.172 & 2.50 & 5.92 $\times 10^{-4}$ & 8.36 $\times 10^{-4}$ & 1.55 $\times 10^{-3}$ & 5.82 $\times 10^{-4}$ \\ 
        0.847 & 1.021 & 2.50 & 4.75 $\times 10^{-4}$ & 6.82 $\times 10^{-4}$ & 1.52 $\times 10^{-3}$ & 4.71 $\times 10^{-4}$ \\ 
        1.320 & 0.852 & 2.50 & 3.55 $\times 10^{-4}$ & 5.23 $\times 10^{-4}$ & 1.54 $\times 10^{-3}$ & 3.59 $\times 10^{-4}$ \\ 
        2.058 & 0.679 & 2.50 & 2.51 $\times 10^{-4}$ & 3.73 $\times 10^{-4}$ & 1.66 $\times 10^{-3}$ & 2.55 $\times 10^{-4}$ \\ 
        3.208 & 0.516 & 2.50 & 1.62 $\times 10^{-4}$ & 2.48 $\times 10^{-4}$ & 1.94 $\times 10^{-3}$ & 1.69 $\times 10^{-4}$ \\ 
        5.000 & 0.379 & 2.50 & 9.99 $\times 10^{-5}$ & 1.56 $\times 10^{-4}$ & 2.43 $\times 10^{-3}$ & 1.06 $\times 10^{-4}$ \\ 
        0.010 & 3.212 & 5.00 & 3.92 $\times 10^{-4}$ & 2.96 $\times 10^{-4}$ & 5.94 $\times 10^{-4}$ & 4.07 $\times 10^{-4}$ \\ 
        0.016 & 3.201 & 5.00 & 3.64 $\times 10^{-4}$ & 2.93 $\times 10^{-4}$ & 5.87 $\times 10^{-4}$ & 4.02 $\times 10^{-4}$ \\ 
        0.024 & 3.179 & 5.00 & 3.57 $\times 10^{-4}$ & 2.91 $\times 10^{-4}$ & 5.77 $\times 10^{-4}$ & 3.95 $\times 10^{-4}$ \\ 
        0.038 & 3.146 & 5.00 & 3.49 $\times 10^{-4}$ & 2.88 $\times 10^{-4}$ & 5.63 $\times 10^{-4}$ & 3.85 $\times 10^{-4}$ \\ 
        0.059 & 3.101 & 5.00 & 3.32 $\times 10^{-4}$ & 2.82 $\times 10^{-4}$ & 5.40 $\times 10^{-4}$ & 3.68 $\times 10^{-4}$ \\ 
        0.092 & 3.031 & 5.00 & 3.38 $\times 10^{-4}$ & 2.73 $\times 10^{-4}$ & 5.08 $\times 10^{-4}$ & 3.45 $\times 10^{-4}$ \\ 
        0.143 & 2.931 & 5.00 & 2.87 $\times 10^{-4}$ & 2.60 $\times 10^{-4}$ & 4.65 $\times 10^{-4}$ & 3.14 $\times 10^{-4}$ \\ 
        0.224 & 2.785 & 5.00 & 2.62 $\times 10^{-4}$ & 2.44 $\times 10^{-4}$ & 4.12 $\times 10^{-4}$ & 2.75 $\times 10^{-4}$ \\ 
        0.349 & 2.594 & 5.00 & 2.39 $\times 10^{-4}$ & 2.20 $\times 10^{-4}$ & 3.47 $\times 10^{-4}$ & 2.31 $\times 10^{-4}$ \\ 
        0.543 & 2.343 & 5.00 & 2.12 $\times 10^{-4}$ & 1.91 $\times 10^{-4}$ & 2.81 $\times 10^{-4}$ & 1.88 $\times 10^{-4}$ \\ 
        0.847 & 2.042 & 5.00 & 1.65 $\times 10^{-4}$ & 1.57 $\times 10^{-4}$ & 2.21 $\times 10^{-4}$ & 1.49 $\times 10^{-4}$ \\ 
        1.320 & 1.705 & 5.00 & 1.19 $\times 10^{-4}$ & 1.21 $\times 10^{-4}$ & 1.73 $\times 10^{-4}$ & 1.13 $\times 10^{-4}$ \\ 
        2.058 & 1.359 & 5.00 & 8.29 $\times 10^{-5}$ & 8.63 $\times 10^{-5}$ & 1.42 $\times 10^{-4}$ & 8.08 $\times 10^{-5}$ \\ 
        3.208 & 1.033 & 5.00 & 5.18 $\times 10^{-5}$ & 5.75 $\times 10^{-5}$ & 1.29 $\times 10^{-4}$ & 5.40 $\times 10^{-5}$ \\ 
        5.000 & 0.759 & 5.00 & 3.17 $\times 10^{-5}$ & 3.62 $\times 10^{-5}$ & 1.28 $\times 10^{-4}$ & 3.41 $\times 10^{-5}$ \\ 
        0.010 & 4.818 & 7.50 & 2.05 $\times 10^{-4}$ & 4.32 $\times 10^{-5}$ & 3.06 $\times 10^{-4}$ & 1.94 $\times 10^{-4}$ \\ 
        0.016 & 4.801 & 7.50 & 2.00 $\times 10^{-4}$ & 4.27 $\times 10^{-5}$ & 3.00 $\times 10^{-4}$ & 1.90 $\times 10^{-4}$ \\ 
        0.024 & 4.769 & 7.50 & 1.95 $\times 10^{-4}$ & 4.26 $\times 10^{-5}$ & 2.91 $\times 10^{-4}$ & 1.84 $\times 10^{-4}$ \\ 
        0.038 & 4.719 & 7.50 & 1.86 $\times 10^{-4}$ & 4.22 $\times 10^{-5}$ & 2.78 $\times 10^{-4}$ & 1.76 $\times 10^{-4}$ \\ 
        0.059 & 4.652 & 7.50 & 1.72 $\times 10^{-4}$ & 4.12 $\times 10^{-5}$ & 2.58 $\times 10^{-4}$ & 1.63 $\times 10^{-4}$ \\ 
        0.092 & 4.546 & 7.50 & 1.54 $\times 10^{-4}$ & 4.00 $\times 10^{-5}$ & 2.32 $\times 10^{-4}$ & 1.47 $\times 10^{-4}$ \\ 
        0.143 & 4.396 & 7.50 & 1.30 $\times 10^{-4}$ & 3.81 $\times 10^{-5}$ & 1.98 $\times 10^{-4}$ & 1.24 $\times 10^{-4}$ \\ 
        0.224 & 4.178 & 7.50 & 1.03 $\times 10^{-4}$ & 3.60 $\times 10^{-5}$ & 1.59 $\times 10^{-4}$ & 9.78 $\times 10^{-5}$ \\ 
        0.349 & 3.892 & 7.50 & 7.27 $\times 10^{-5}$ & 3.25 $\times 10^{-5}$ & 1.17 $\times 10^{-4}$ & 6.96 $\times 10^{-5}$ \\ 
        0.543 & 3.515 & 7.50 & 4.59 $\times 10^{-5}$ & 2.85 $\times 10^{-5}$ & 7.88 $\times 10^{-5}$ & 4.55 $\times 10^{-5}$ \\ 
        0.847 & 3.063 & 7.50 & 2.65 $\times 10^{-5}$ & 2.34 $\times 10^{-5}$ & 4.94 $\times 10^{-5}$ & 2.87 $\times 10^{-5}$ \\ 
        1.320 & 2.557 & 7.50 & 1.88 $\times 10^{-5}$ & 1.81 $\times 10^{-5}$ & 3.00 $\times 10^{-5}$ & 1.90 $\times 10^{-5}$ \\ 
        2.058 & 2.038 & 7.50 & 1.42 $\times 10^{-5}$ & 1.30 $\times 10^{-5}$ & 1.88 $\times 10^{-5}$ & 1.29 $\times 10^{-5}$ \\ 
        3.208 & 1.549 & 7.50 & 8.65 $\times 10^{-6}$ & 8.67 $\times 10^{-6}$ & 1.33 $\times 10^{-5}$ & 8.61 $\times 10^{-6}$ \\ 
        5.000 & 1.138 & 7.50 & 5.35 $\times 10^{-6}$ & 5.46 $\times 10^{-6}$ & 1.03 $\times 10^{-5}$ & 5.48 $\times 10^{-6}$ \\ \hline 
        \caption{Model grid and mass-loss rates. We consider $\lambda_p = 2.5, 5.0,$ and 7.5, and and vary the Roche-lobe filling parameter $f_\phi$ between 0.01 and 5 (15 logarithmically spaced values). For each pair, the corresponding $\lambda_{\rm Hill}$ is computed. The table lists $f_\phi$, $\lambda_p$, $\lambda_{\rm Hill}$, the simulated mass-loss rate ($\dot{M}_{\rm sim}$, and predictions from the Parker-wind ($\dot{M}_{\rm PW}$), nozzle ($\dot{M}_{\rm nozzle}$), and mixture ($\dot{M}_{\rm mix}$) models.}
        \label{tab:models}
    \end{longtable}
\end{table}

\subsection{Planetary Boundary Conditions}
The effective gravitational potential in the frame that corotates with the orbit at a given point, $(x,y,z)$ is defined as
\begin{equation}
    \Phi_{\rm eff} = -\frac{GM_\star}{|\boldsymbol{r}_\star|} -\frac{GM_p}{|\boldsymbol{r}_p|} -\frac{1}{2}\Omega^2R^2,
\end{equation}
\\
where $R = \sqrt{x^2 + y^2}$. In our model, the planetary surface is treated as a fixed boundary condition defined by a critical potential, $\Phi_{s}$. The modeled planet thus corotates with the orbital frequency. In regions close to the planet where $\Phi_{\rm eff} < \Phi_{s}$ we set the gas velocities to zero in the corotating frame of the calculation and remove all acceleration terms. The surface density ($\rho_s$) and pressure ($P_s$) of the planet are set to fixed values. Since we neglect the self-gravity of the gas, we set $\rho_s = M/a^3 = 1$, without loss of generality. The surface pressure is determined by the hydrodynamic escape parameter at the Hill sphere,
\begin{equation}
    P_s = \frac{\rho_s c_s^2}{\gamma} = \frac{\rho_s |\Phi_{\rm Hill}|}{\gamma \lambda_{\rm Hill}}.
\end{equation}
We specify $\Phi_s$ using the effective potential at the $L_1$ Lagrange point as
\begin{equation}
    \Phi_s = \Phi_{\rm eff}(L_1) + f_{\phi} \Phi_{\rm Hill}
\end{equation}
where $f_{\phi}$ is a dimensionless scaling factor that can be varied between models. Smaller values of $f_{\phi}$ correspond to more Roche-lobe-filling planets, while larger values represent more spherical planets that are deeper within their Roche lobes and thus less affected by the tidal potential. 

As stated above, the value of $\Phi_s$ defines the planetary surface boundary and, hence, the volume enclosed by the planet. The total planetary volume $V_p$ is approximated by integrating (i.e., summing over the discrete mesh) the volumes of all grid cells that lie within the planetary surface. Because this volume is tidally distorted by the Roche potential, we define an effective planetary radius, $R_{\rm eff} = (3V_p/4\pi)^{1/3}$. Computing $R_{\rm eff}$ allows us to relate $\lambda_p$ and $\lambda_{\rm Hill}$ by equating the gas sound speed, 
\begin{equation} \label{eq: lambda relation}
    \lambda_p = \frac{r_{\rm H} \lambda_{\rm Hill}}{R_p},
\end{equation}
where we take $R_p \approx R_{\rm eff}$. 

\begin{figure*}
    \centering
    \includegraphics[width=\linewidth]{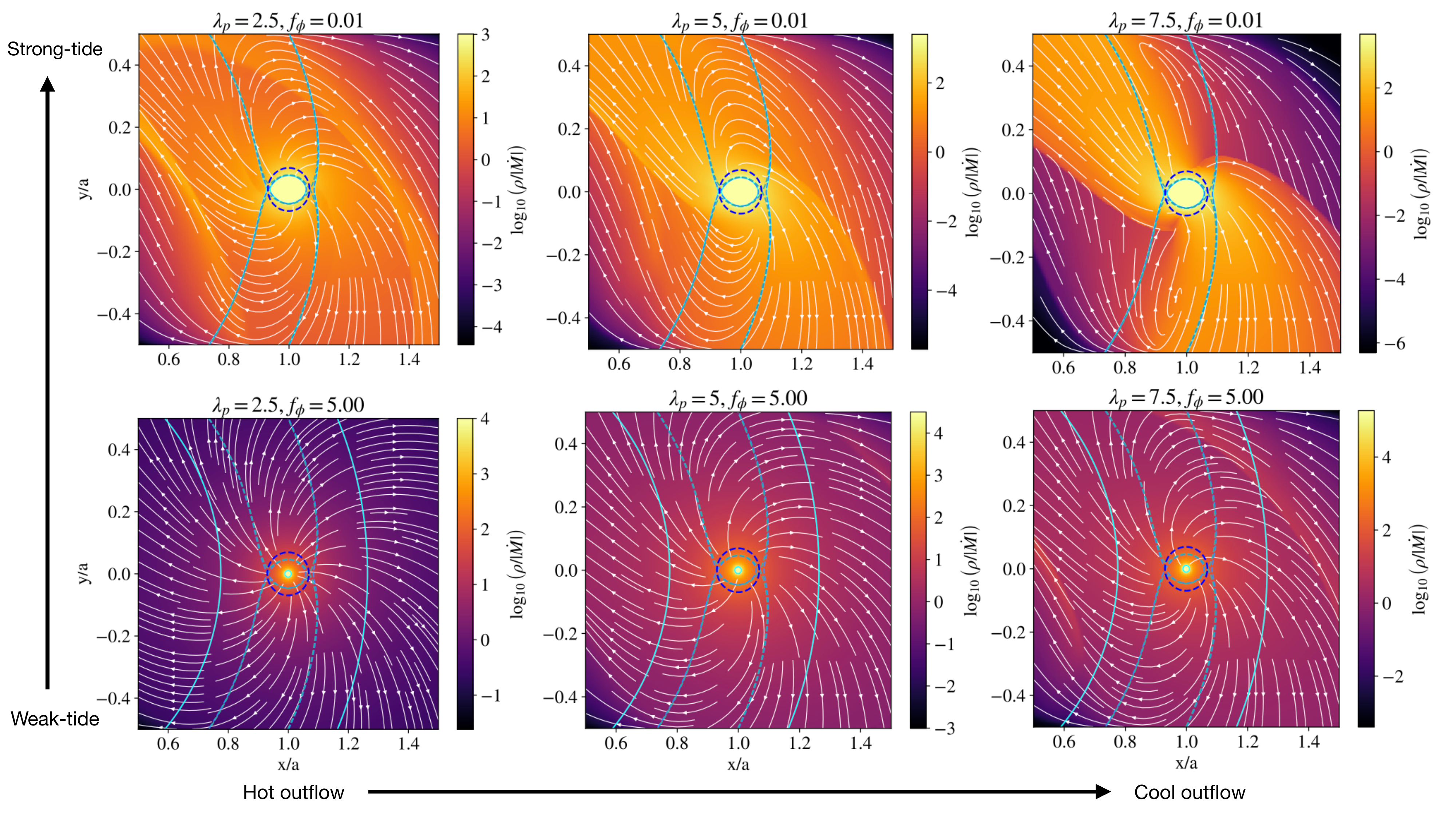}
    \caption{Wind morphology in six representative models spanning $(f_\phi, \lambda_p)$. Shown are slices of the gas density in the star–planet equatorial plane, normalized by the total mass-loss rate, for three values of the $\lambda_p$	(columns) and two degrees of Roche-lobe filling, $f_\phi$ (rows). The top row ($f_\phi = 0.01)$ corresponds to a more Roche-lobe-filling configuration, while the bottom row ($f_\phi = 5.00$) shows a planet deeper within its Roche lobe. Increasing $\lambda_p$ corresponds to cooler, more tightly bound atmospheres. White arrows show streamlines in the corotating frame. The solid cyan curve denotes the $\Phi_s$ contour, and the dashed contour lines correspond to the potential at the Lagrange points. The dark blue circle marks the intersection of the Hill sphere with the equatorial-plane slice. The figure illustrates the transition from a spherical outflow in the embedded, hot-wind regime to a two–tidal-tail morphology as the planet approaches Roche-lobe filling and as the atmosphere has less thermal energy.}
    \label{fig:wind_morphology}
\end{figure*}

\subsubsection{Computational Domain}
The simulation domain extends from $0.5a$ to $1.5a$ in the $x-$direction, with the planet centered approximately at $x \simeq a$, and spans $\pm 0.5a$ in both the $y-$ and $z-$directions. The base mesh consists of $256^3$ zones. We include one additional level of static mesh refinement surrounding the planet, covering the region $0.75a \leq x \leq 1.25a$ and $\pm0.25a$ in the $y-$ and $z-$ directions containing $256^3$ refined zones. This doubles the resolution in each direction within the refined region. The smallest zones are therefore cubes with sides $0.5/256 \approx 0.00195$. The mesh is decomposed into meshblocks of $16^3$ zones each. Our grid resolves the smallest effective planet radius in the simulation, $R_{\rm eff} \simeq 1.05 \times 10^{-2}$, by $\sim 11$ zones across the diameter. We also performed resolution convergence tests which show that $\dot{M}$ is approximately converged at a coarser resolution than our grid, indicating that the measured mass-loss rates are not dominated by grid-scale effects (refer to Appendix \ref{app:resolution} for more details).

The outer boundary condition in each direction is an outflow-only “diode” boundary where gas is permitted to exit the domain but is not allowed to re-enter it. Velocities normal to the $+x/+y/+z$ boundaries are restricted to be positive, while those normal to the $-x/-y/-z$ boundaries are restricted to be negative. This ensures that material can leave the computational domain freely but can not flow back into it from any boundary. These diode conditions are implemented using user–defined boundary functions in Athena++.

\subsection{Mass Loss Rate Diagnostic}
We measure the planetary mass-loss rate by evaluating the flux of material flowing outward through a closed surface, $S$, surrounding the planet via a surface integral,
\begin{equation}
    \dot{M} = -\oint_S \rho(\boldsymbol{\nu} \cdot \boldsymbol{dA}),  
\end{equation}
where $\boldsymbol{dA}$ is the outward-facing area element on $S$. In practice, we compute this surface integral by summing over all zones that lie closest to a spherical surface of a given radius, centered on the planet. We adopt a spherical surface of radius $0.3a$, but we have verified that the inferred $\dot{M}$ remains nearly unchanged for any choice of radii in the range $0.2a-0.5a$. 

\subsection{Models}
We run a suite of models varying $\lambda_p$ and $f_{\phi}$. We set $\lambda_p = 2.5, 5.0$, and 7.5 and for each value of $\lambda_p$, we vary $f_\phi$ between 0.01 and 5 (15 logarithmically spaced values). Our parameter choices (Table \ref{tab:models}) are intended to span a realistic range of combinations relevant to hot gas giant planets. For every pair ($\lambda_p, f_\phi$), the corresponding $\lambda_{\rm Hill}$ is calculated as described above in Eq.\ref{eq: lambda relation}. In the following sections, we explore these 45 combinations of $\lambda_p, \lambda_{\rm Hill}$, and $f_{\phi}$. 

\section{Morphology and Kinematics of the 3D Outflow} \label{sec-results}
\subsection{Wind Morphology} \label{sec-wind morphology}

The thermal winds in our model are launched from rest at the corotating planetary surface. Figure \ref{fig:wind_morphology} shows slices of the wind density in the star–planet equatorial plane, normalized by the mass-loss rate and focusing on the region near the planet for six representative cases. Flow streamlines in the corotating frame are overplotted. The figure shows that the wind density structure in the vicinity of the planet depends strongly on the degree of Roche-lobe filling, parameterized by $f_\phi$, and on the hydrodynamic escape parameter $\lambda_p$, which measures how tightly bound the atmosphere is relative to its thermal energy.

In Figure \ref{fig:wind_morphology}, the top row corresponds to a more Roche-lobe-filling case ($f_\phi = 0.01$, the smallest we simulated), while the bottom row shows planets embedded deeper within their Roche-lobes ($f_\phi = 5.00$, the largest we simulated). For fixed $\lambda_p$, varying $f_\phi$ can be interpreted as placing the same planet at different orbital separations. Smaller $f_\phi$ corresponds to a closer-in configuration where the Roche geometry is more pronounced and the atmosphere lies nearer to Roche-lobe overflow, whereas larger $f_\phi$ corresponds to a wider separation where tidal effects are weaker and the planet remains spherical and more deeply embedded within its Roche lobe. 

\begin{figure*}
    \centering
    \includegraphics[width = \linewidth]{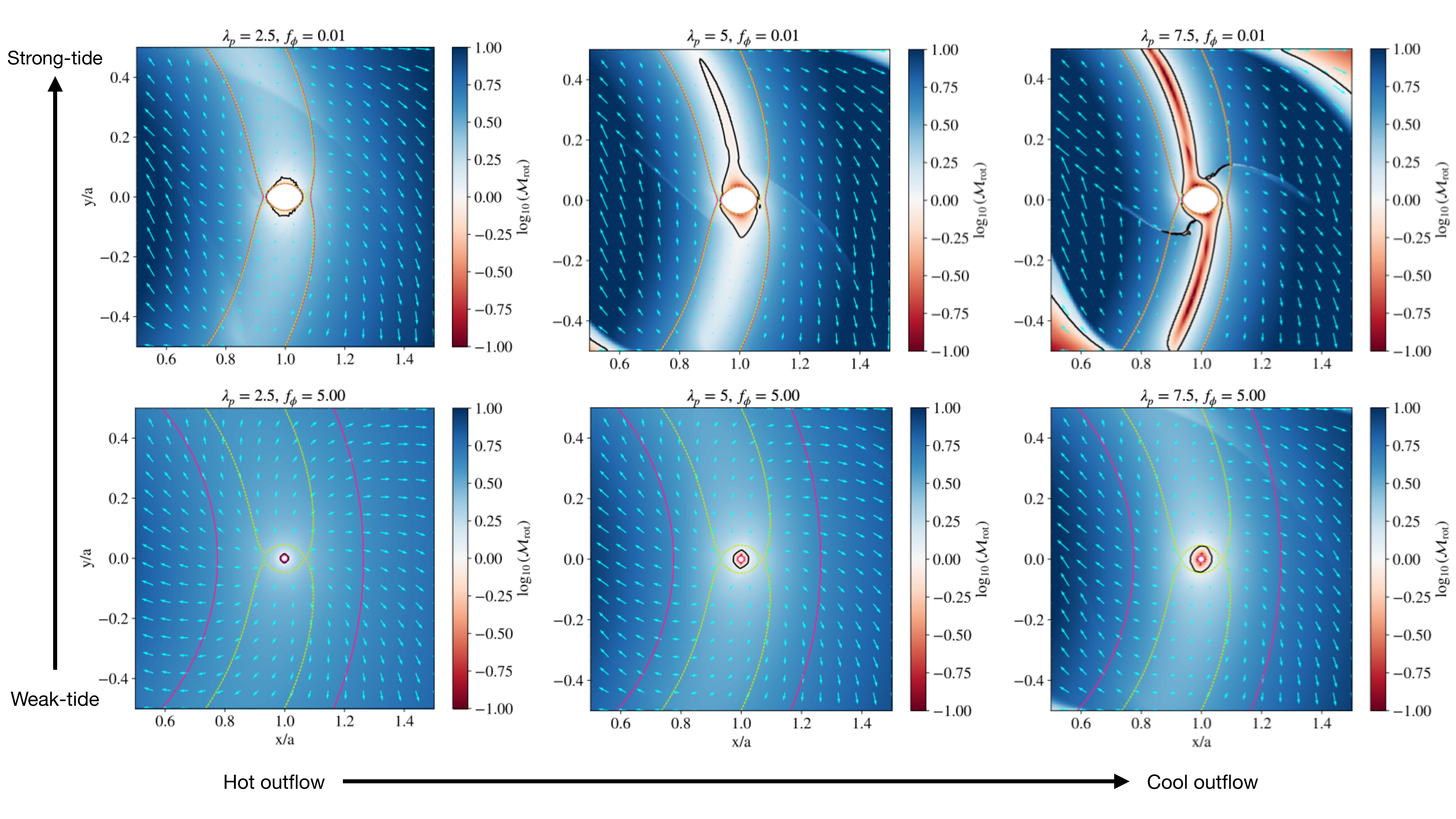}
    \caption{Star-planet equatorial plane slices for the same six models shown in Figure \ref{fig:wind_morphology}, color-coded by the Mach number in the corotating frame. Arrows indicate the velocity field in the corotating frame, and the black contour marks the sonic surface ($\cal M_{\rm rot}$ = 1). The magenta contour shows $\Phi_s$, and the green and gold contours denote the effective potential contours passing through the $L_1$ and $L_2$	points, respectively.}
    \label{fig:mach_number}
\end{figure*}

The columns correspond to $\lambda_p = 2.5$ (Figure \ref{fig:wind_morphology}, Left),  $\lambda_p = 5$ (Middle), and $\lambda_p = 7.5$ (Right). Since $\lambda_p$ is inversely related to the Parker wind temperature at the launch boundary ($c_s^2 \propto T$), increasing $\lambda_p$ corresponds to cooler winds and a more tightly bound atmosphere, whereas smaller $\lambda_p$ corresponds to hotter winds for which thermal energy more readily overcomes the planet’s gravitational potential, facilitating atmospheric escape.
For the model with $f_\phi = 0.01$ and $\lambda_p = 2.5$ (Figure \ref{fig:wind_morphology}, top left), the high speed at which the outflow is launched produces a rapidly accelerating outflow that becomes supersonic close to the planet (discussed in more detail in \S \ref{sec:wind acc and velocity}). In this hot-wind regime, the gas escapes efficiently over a broad range of solid angles, producing a nearly spherical gas density distribution near the planet that becomes increasingly tidally channeled at larger distances due to tidal effects. By contrast, for $f_\phi = 0.01$ and $\lambda_p = 7.5$ (Figure \ref{fig:wind_morphology}, top right), the tidal potential is the same but the cooler wind is more gravitationally bound and cannot escape as readily in all directions. Instead, the outflow is preferentially funneled along the stellar and anti-stellar directions, with most material escaping through the vicinity of the $L_1$ and $L_2$ Lagrange points and forming a two–tidal-tail morphology. These tails follow curved, non-radial trajectories, due to the effects of the Roche potential and rotational (Coriolis) forces. 

Models with same $f_\phi$ and intermediate $\lambda_p$ show a mixed morphology where the flow is roughly spherical very close to the planetary surface, but at larger distances it becomes increasingly collimated into tidal tails. The distance from the planet at which this transition occurs decreases as tidal effects strengthen, i.e., for more Roche-lobe-filling and more tightly bound (cooler) wind configurations. We note that these simulations model thermally driven, collisional hydrodynamic escape. In real systems, non-thermal escape processes and interactions with the stellar wind can also fill the Roche lobe and modify the resulting outflow morphology \citep{2020JGRA..12527639G}, even for magnetized planets \citep{2018A&A...614L...3G}. These effects are not included here and should be explored in future work.

We find that the planetary outflow morphology varies smoothly across ($f_\phi, \lambda_p$). The most pronounced two–tidal-tail structures occur when the tidal influence is strong (small $f_\phi$, i.e., near Roche-lobe filling) combined with cooler, more tightly bound atmospheres (larger $\lambda_p$). In contrast, when the tidal influence is weaker (large $f_\phi$, i.e., the planet is deeply embedded within its Roche lobe) together with hotter winds (smaller $\lambda_p$) a spherical, planet-centered outflow approaching the isolated planet limit is favored. 

This morphological transition is complementary to the “bubbles and streams” picture described by \citet{2025ApJ...988...63M}, who modeled planetary outflows interacting with an ambient stellar wind. In our simulations, hot, weakly bound winds escape over a broad solid angles, once embedded in a stellar-wind environment, such flows naturally connect to extended, bubble-like morphologies. Cooler and more tightly bound winds, on the other hand, are intrinsically anisotropic, with escape preferentially channeled through the vicinity of $L_1$ and $L_2$, analogous to stream-like outflows. Thus, although we do not model the stellar-wind interaction, our simulations isolate the tidal and thermal conditions that set the underlying launch geometry of bubble-like versus stream-like planetary winds.
\begin{figure*}
    \centering
    \includegraphics[width=\linewidth]{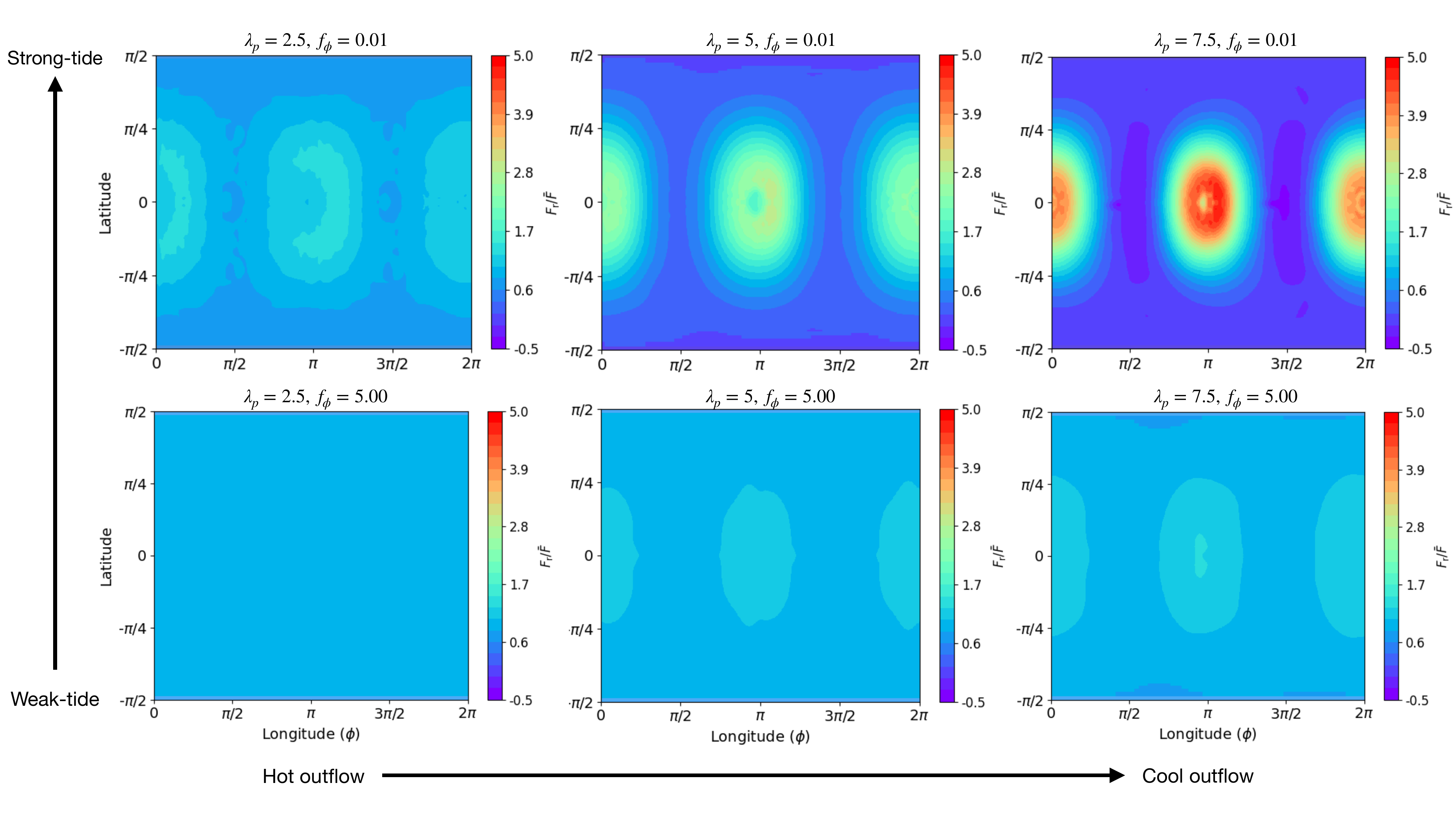}
    \caption{Angular distribution of the normalized radial mass flux on the Hill surface for the same six models shown in Figures \ref{fig:wind_morphology} \& \ref{fig:mach_number}. Each panel shows a latitude–longitude map of $Z(\theta,\phi) \equiv F_r/\bar{F}$, evaluated on the Hill surface. Longitude, $\phi$ is measured in the orbital plane; the $L_1$ and $L_2$ points lie at latitude $0^\circ$, at 
    $\phi = \pi$, and $\phi = 0$ (or $2\pi$), respectively.}
    \label{fig:radial profile}
\end{figure*}

\subsection{Wind Acceleration and Velocity} \label{sec:wind acc and velocity}
Thermal winds accelerate due to radial pressure gradients, starting from rest on the planetary surface and passing through a critical point where the expansion speed equals the local sound speed, before becoming supersonic. In Figure \ref{fig:mach_number}, we show slices through the planet’s orbital plane of the Mach number in the corotating frame. The Mach number $\cal M_{\rm rot}$ is defined as the ratio of the flow speed in the corotating frame to the local sound speed. The velocity vectors (in the corotating frame) are overplotted as a quiver map. The $\cal M_{\rm rot}$ = 1 contour (in black) marks the sonic surface. 

For an isothermal spherical Parker wind in the isolated-planet limit, the sonic radius is $r_{\rm sonic} = GM_p/(2c_s^2)$, which is equivalent to $r_{s} = R_p(\lambda_p/2)$ \citep{1999isw..book.....L}. This scaling implies a larger subsonic region for larger $\lambda_p$, as cooler winds have weaker pressure support and accelerate more slowly. In Roche geometry, however, the effective potential is non-spherical and the flow comprises a quasi-spherical component near the planet and tidally directed two-tail outflows at larger distances (see \S \ref{sec-wind morphology}). As a result, the $\cal M_{\rm rot} = $ 1 surface or (sonic surface) is not expected to be perfectly spherical either. 

In the weak-tide, less Roche-lobe filling regime (large $f_\phi$; bottom row of Figure \ref{fig:mach_number}), the case that most closely approaches the isolated-planet limit -- $f_\phi = 5.00$ with the hottest wind ($\lambda_p = 2.5$) -- shows a sonic surface ($\cal M_{\rm rot} = $ 1) that is nearly spherical and lies very close to the planetary surface, indicating rapid acceleration to supersonic speeds shortly after launch. As $\lambda_p$ increases at the same $f_\phi$, the sonic surface moves outward, consistent with the behavior predicted by the spherical Parker-wind solution, in which cooler, more tightly bound atmospheres exhibit a larger subsonic region. We also find that the sonic surface becomes mildly non-spherical at larger $\lambda_p$	(bottom middle and bottom right), developing a slight inward distortion along the star-plant line. This suggests that even in the weak-tide regime, residual tidal effects become more apparent for cooler, more tightly bound winds, leading to preferential acceleration toward the $L_1/L_2$ directions relative to directions perpendicular to the star–planet axis.

In the strong-tide, more Roche-lobe filling regime (small $f_\phi$; top row), departures from spherical symmetry are more pronounced. Even for $\lambda_p = 2.5$, the sonic surface shows some distortion, with a slight extension along directions perpendicular to the star–planet axis, although the flow still reaches supersonic speeds relatively close to the planet in all directions. As $\lambda_p$ increases, the sonic surface expands and becomes increasingly anisotropic: the gas reaches $\cal M_{\rm rot}$ = 1 more readily along the star–planet axis (i.e., toward the $L_1/L_2$ directions), while remaining subsonic to larger radii in perpendicular directions, and in the most extreme case ($\lambda_p = 7.5$) extending beyond the simulated domain. This trend is consistent with the outflow morphology discussed in \S \ref{sec-wind morphology}, where cooler, more tightly bound winds are increasingly shaped by the Roche potential and escape preferentially through the vicinity of the Lagrange points rather than quasi-spherically. 

Finally, we note two other asymmetries in the sonic surface around the planet. First, there is an asymmetry along the star–planet axis. This is most pronounced in the strongly tidally influenced, cooler-wind case (Figure \ref{fig:mach_number} top right), where the flow reaches the sonic surface at smaller radii on the stellar ($L_1$) side than on the anti-stellar ($L_2$) side, indicating reduced acceleration (and hence outflow efficiency) toward $L_2$. A contributing factor is rotation in the corotating frame, where Coriolis forces deflect the flow, producing non-radial streamlines that differentially modify streamline curvature and acceleration on the two sides. Second, the sonic surface can be asymmetric in the direction perpendicular to the star–planet line: the transition to $\cal M_{\rm rot}$ = 1 occurs at smaller radii on $-y$ side of the orbital-plane slice than on the other $+y$ side, which remains subsonic to larger distances. Because the Roche potential is symmetric under $+y \to -y$, this perpendicular asymmetry is also consistent with rotational effects in the corotating frame (e.g., Coriolis deflection), which naturally introduce a leading–trailing difference in the streamline geometry and acceleration.

\subsection{Angular Mass-Flux Profile on the Hill Sphere} 
\label{sec:radial profile}
While Figures \ref{fig:wind_morphology} \& \ref{fig:mach_number} illustrate the anisotropy in the equatorial plane, we further characterize its full 3D structure by computing the radial mass flux through a spherical surface of radius $r = r_{\rm H}$ centered on the planet. For a given model, at every angular location $(r_{\rm H}, \theta, \phi)$ on the Hill surface, we compute the radial flux density, $F_r(r_{\rm H}, \theta, \phi) = \rho(r_{\rm H}, \theta, \phi)[\boldsymbol{v}(r_{\rm H}, \theta, \phi)\cdot\hat{\boldsymbol{r}}]$, where $\hat{\boldsymbol{r}}$ is the outward radial unit vector from the planet, and $\rho$ and $\boldsymbol{v}$ 
are the gas density and velocity interpolated onto the surface of the Hill sphere. We then normalize by the surface-averaged radial flux,
\[
\bar{F} \equiv \frac{\dot{M}(r_{\rm H})}{4\pi r_{\rm H}^2},
\]
where $\dot{M}(r_{\rm H})$ is the total mass-loss rate through the Hill surface given by, $\dot{M}(r_{\rm H}) = \oint_{S(r_{\rm H})} \rho\,(\mathbf{v}\cdot\hat{\mathbf r})\, dA$, and we define the dimensionless angular flux distribution as,
\[
Z(\theta,\phi) \equiv \frac{F_r(\theta,\phi)}{\bar{F}}.
\]
$Z(\theta,\phi) \simeq 1$ at all angular locations corresponds to nearly isotropic escape, whereas significant angular variations, i.e., $Z \ll 1$ in some directions and $Z \gg 1$ in others indicate strongly anisotropic, collimated outflow. Negative values, $Z<0$ correspond to localized inflow/recirculation across the Hill surface. 

In Figure~\ref{fig:radial profile}, we show latitude–longitude maps of the normalized radial flux, $Z(\theta, \phi)$, evaluated on the Hill sphere for six representative models shown in Figures \ref{fig:wind_morphology} \& \ref{fig:mach_number}. Here, latitude is defined as $\pi/2 - \theta$ and longitude is equivalent to the azimuthal angle, $\phi$.

In the weak-tide (less Roche lobe filling) and hot-wind regime (large $f_\phi$, small $\lambda_p$), the map $Z(\theta, \phi)$ is nearly uniform, indicating that the radial flux through the Hill sphere varies weakly with angle and that escape is approximately spherical (Figure \ref{fig:radial profile}, bottom left). In contrast, in the strong-tide and cool-wind regime (small $f_\phi$, large $\lambda_p$), the flux becomes highly anisotropic and concentrates into two narrow lobes in the orbital plane aligned with the star–planet axis. These maxima occur near longitude, $\phi = \pi$ and $\phi = 0/2\pi$  (latitude $0^\circ$), corresponding to the stellar and anti-stellar directions, respectively, consistent with preferential escape through the vicinity of the $L_1$ and $L_2$ Lagrange points. Intermediate $(f_\phi, \lambda_p)$ cases show a smooth transition, with a quasi-spherical component near the planet and increasing angular concentration toward the $L_1/L_2$ directions as tidal shaping strengthens and/or the atmosphere becomes more tightly bound. These results are consistent with the morphological trends inferred in \S \ref{sec-wind morphology} \& \ref{sec:wind acc and velocity} from the density and Mach-number maps in the equatorial plane.

\section{Modeling Mass-Loss Rates} \label{sec:mass_loss rate models}
Our goal is to derive a closed-form analytical expression for the mass-loss rate by combining insights from our 3D hydrodynamic simulations with the physical picture of how gas escapes the planet as described above.

\subsection{Spherical Outflow Model} \label{sec:sph outflow formula}
\begin{figure*}
    \centering
    \includegraphics[width=\linewidth]{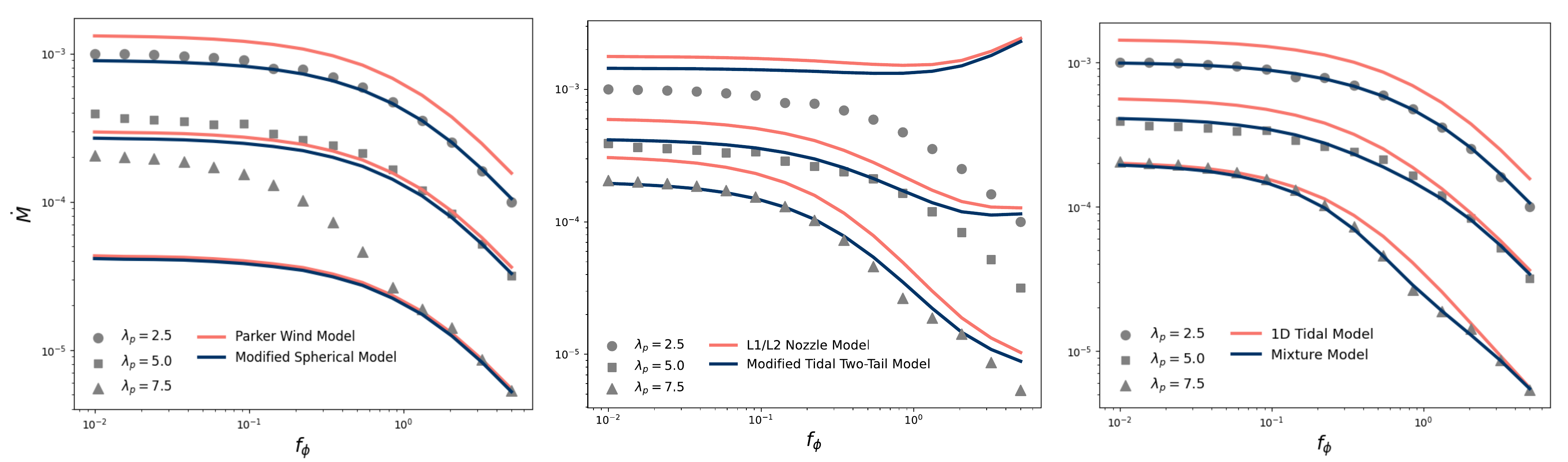}
    \caption{Figure comparing the mass-loss rates predicted by the analytic models described in \S \ref{sec:mass_loss rate models} with those measured from our 3D hydrodynamic simulations, which are shown by the grey markers.}
    \label{fig:models_comparison}
\end{figure*}
In the exoplanet literature, mass-loss rates are often estimated using energy-limited photoevaporation models, in which a fraction of the incident high-energy stellar flux is converted into work against the planet’s gravitational potential to drive atmospheric escape \citep{1981Icar...48..150W, 2017ApJ...847...29O, 2019ApJ...874...91W}. However, such models do not describe the dynamical structure of the outflow. The Parker wind (PW) \citep{1960ApJ...132..821P}, in contrast, provides a hydrodynamic solution for a thermally driven, transonic flow given a specified thermal state (or sound speed), and is therefore commonly used as a physically motivated baseline for estimating mass-loss rates. However, its regime of validity, particularly in the presence of strong tidal effects, is not always well characterized.
The classic Parker solution is strictly isothermal; however, our simulations adopt a nearly isothermal equation of state ($\gamma=1.01$), so we use the isothermal Parker wind model as an approximate benchmark. In this limit, the outflow is driven by thermal pressure, passes through a sonic point, and approaches a steady radial wind with mass-loss rate given by \citep{1999isw..book.....L},
\begin{equation} \label{eq:PW}
    \dot M_{\rm PW} = \pi\rho_s\sqrt{GM_pR_p^3\lambda_p^3} \exp\left( \frac{\Delta\Phi}{c_s^2} - \frac{1}{2} \right),  
\end{equation}
where $\Delta\Phi$ is the potential depth that must be overcome by the gas for the escape. Here, we take ${\Delta\Phi=\Phi_s-\Phi(r_{\rm s})}$, the difference between the gravitational potential at the planetary surface and at the sonic radius, $r_{\rm s}$. The PW model assumes a spherically symmetric outflow, such that the sonic surface is a sphere of radius $r_s$.
We note, $\Phi_s/c_s^2 = -\lambda_p$, and for an isothermal wind, the sonic radius satisfies $r_{\rm s} = GM_p/(2c_s^2)$, giving  $-\Phi(r_{\rm s})/c_s^2 = 2$. Applying this, Eq. \ref{eq:PW} is reduced to,
\begin{equation} \label{eq:sonic PW}
    \dot{M}_{\rm PW} = \pi\rho_s\sqrt{GM_pR_p^3\lambda_p^3} \exp\left( \frac{3}{2} - \lambda_p \right).
\end{equation}

We now compare the spherical Parker-wind model to our simulation results (Figure \ref{fig:models_comparison} (Left); red curves). Based on the discussions in \S \ref{sec-wind morphology}, \S \ref{sec:wind acc and velocity} \& \S\ref{sec:radial profile}, we expect this model to perform best in the regime that closely approaches the isolated-planet limit, i.e., large $f_\phi$ (weak tidal influence / low Roche-lobe filling) and small $\lambda_p$ where the flow morphology remains approximately spherical. However, we find that at small $\lambda_p$ (hotter winds) the Parker-wind model captures the overall trend but tends to overpredict the mass-loss rates. In this regime, the sonic point lies very close to the planetary surface (see Figure \ref{fig:mach_number}), so the mass-loss rate becomes sensitive to the detailed structure of the flow near the base. 

To account for this, we introduce an empirical correction factor, $g(\lambda_p)$ to Eq. \ref{eq:sonic PW} and define a \textit{Modified Spherical Model} to estimate mass-loss rates, $M_{\rm sph}$ given by
\begin{equation} \label{eq:msphere}
    \dot M_{\rm sph} = g(\lambda_p)\dot M_{\rm PW},
\end{equation}
where we adopt $g(\lambda_p) = \exp(-k_{\rm sph}/\lambda_p^2)$. Here 
$k_{\rm sph}$ is a constant calibrated by fitting $\dot M_{\rm sph}$
to the simulation obtained mass-loss rates in the regime where the flow remains quasi-spherical. We find $k_{\rm sph} = 2.47$. 

Figure \ref{fig:models_comparison}, Left shows that the \textit{Modified Spherical Model} provides a good approximation in the weak-tide, less Roche-lobe-filling regime, where the outflow remains nearly isotropic (large $f_\phi$, low $\lambda_p$). For hot, loosely bound atmospheres ($\lambda_p = 2.5$), the \textit{Modified Spherical Model} reproduces the mass-loss rates from 3D simulations across the full range of degrees of Roche-lobe filling or $f_\phi$ values explored. For more tightly bound atmospheres ($\lambda_p = 5,$ 7.5), good agreement is largely limited to systems with $f_\phi \gtrsim 1$. However, at lower $f_\phi$, the magnitude of the discrepancy depends on the wind temperature and the resulting degree of anisotropy. For the intermediate case ($\lambda_p = 5$), the model underpredicts the mass-loss rates but still captures the correct order of magnitude, while for cooler winds ($\lambda_p = 7.5$), the deviation from the 3D simulations can approach an order of magnitude. The \textit{Modified Spherical Model} breaks down in this regime as the assumption of spherical symmetry becomes invalid and the outflow geometry is more strongly shaped by tidal forces.

\subsection{Tidal Two-Tail Model} \label{sec:tidal tail outflow formula}
As shown in \S \ref{sec-wind morphology} \& \S \ref{sec:radial profile}, the outflow becomes highly anisotropic and develops a two-tail morphology, in more Roche-lobe filing, strong tide, and cool winds regime where the spherical models (Eq.\ref{eq:PW} \& \ref{eq:msphere}) no longer capture the effective escape geometry.

In this limit, the gas preferentially escapes along the star-planet axis through the vicinity of the $L_1$ and $L_2$ Lagrange points (see Figure \ref{fig:radial profile}). As seen in \S\ref{sec:wind acc and velocity}, the sonic surface is reached much closer to the planet along these directions, while remaining extended in the transverse directions, reflecting a suppressed lateral escape. As a result, the total mass-loss rate is dominated by the $L_1/L_2$ channels, motivating a simplified “\textit{nozzle}” approximation in which the atmospheric escape is restricted through these nozzles. Following \citet{1975ApJ...198..383L, 2017ApJ...835..145J}, we derive the nozzle framework where we treat the flow as steady and nearly isothermal along the star–planet axis, with transverse hydrostatic balance set by the local curvature of the effective potential, $\Phi_{\rm eff}$. The effective cross-section of each stream is obtained by the quadratic (second-order) expansion of $\Phi_{\rm eff}$ about the corresponding Lagrange point, retaining only the leading curvature terms in directions perpendicular to the flow. The resulting mass-loss rate, $\dot{M}_{\rm noz}$, by the \textit{Nozzle Model} is given by:
\begin{equation}
    \dot M_{\rm noz} = \dot M_{L_1} + \dot M_{L_2}
\end{equation}
where $\dot M_{L_1}$ and $\dot M_{L_2}$ denote the contributions through the $L_1$ and $L_2$ channels, respectively and are computed using:
\begin{equation}
\begin{aligned}
\dot M_{L_i} \approx\;& \rho_s c_s^3
\left(\frac{2\pi}{\sqrt{\phi_{yy}^i \phi_{zz}^i}}\right) \exp\!\left[\frac{\Phi_s- \Phi_{\rm eff}(L_i)}{c_s^2}
-\frac{1}{2}\right].
\end{aligned}
\end{equation}
for $i = 1, 2$. Here $\Phi_{\rm eff}(L_i)$ is the effective potential at $L_i$ Lagrange point. The curvature terms are defined as: 
\[
\phi_{yy}^{\,i} \equiv \left.\frac{\partial^2 \Phi_{\rm eff}}{\partial y^2}\right|_{L_i},
\qquad
\phi_{zz}^{\,i} \equiv \left.\frac{\partial^2 \Phi_{\rm eff}}{\partial z^2}\right|_{L_i},
\quad (i=1,2).
\]
and are evaluated at the corresponding Lagrange points. For our setup, we obtain $\phi_{yy}^1=3.4538$, $\phi_{zz}^1=4.4538$, $\phi_{yy}^2=2.6173$ and $\phi_{zz}^2=3.6173$. While fixed within our simulations due to the adopted normalization ($G=M=a=1$) and mass ratio, these values are not universal and depend on the system parameters, but can be easily computed for any given configuration from the effective potential. 

Since the model assumes that mass loss is dominated by the $L_1/L_2$ funnels, it is expected to be valid primarily in the strong-tide, cool wind regime (small $f_\phi$, large $\lambda_p$), where the outflow exhibits a two-tail, anisotropic structure and the underlying assumptions are approximately satisfied. However, a comparison with our 3D simulations shows that the model systematically overpredicts the mass-loss rate, even in this regime (Figure~\ref{fig:models_comparison}, middle).

To identify the source of this overprediction, we revisit the sonic surface maps discussed in \S\ref{sec:wind acc and velocity}. In the high-tide, cool-wind regime (Figure~\ref{fig:mach_number}, top right), the simulations reveal a modest asymmetry in the location of the sonic surface along the star–planet axis, with the transition occurring at slightly larger radii on the anti-stellar ($L_2$) side and at smaller radii on the star-facing ($L_1$) side. This is also the regime in which the nozzle approximation would otherwise be expected to perform well. Consistently, the corresponding latitude–longitude maps (Figure~\ref{fig:radial profile}, top right) show that the normalized mass flux, $Z(\theta,\phi)$, is slightly enhanced toward $L_1$ ($\phi = \pi$) relative to $L_2$ ($\phi = 0, 2\pi$).

The asymmetry observed in the high-tide, cool-wind regime likely arises from the interplay between the effective potential and rotating-frame dynamics. Compared to the $L_1$ direction, the potential gradient toward $L_2$ is shallower, leading to slower acceleration and a more extended subsonic region. As a result, gas flowing toward $L_2$ travels farther in the accelerating region, making it more susceptible to Coriolis deflection, whereas the $L_1$ flow reaches supersonic speeds at smaller radii and is therefore less affected by rotational effects. This naturally leads to some suppression of mass flux through the $L_2$ channel.
The importance of rotational effects can be quantified by the dimensionless Rossby number ($\rm Ro$) given by,
\begin{equation}
\mathrm{Ro} = \frac{c_s}{\Omega \ell} = \sqrt{\frac{G M_p}{R_p \lambda_p}}\frac{1}{\Omega\ell},
\end{equation}
where we adopt $\ell = |x_{L_2} - x_p|$ as the characteristic length scale of the $L_2$ funnel, and $x_{L_2}$ is the x-coordinate of the $L_2$ point. For our setup $x_{\rm L_2} = 1.0699$. $\rm Ro$ is defined as a dimensionless ratio that compares the characteristic outflow speed to orbital advection in the rotating frame. Since $\mathrm{Ro} \propto c_s$, cooler, more tightly bound winds (larger $\lambda_p$) also correspond to a low-$\mathrm{Ro}$ limit, in which Coriolis forces become more important. While the ratio of Coriolis to inertial acceleration scales as $\mathrm{Ro}^{-1}$, the resulting reduction in mass flux arises through streamline deflection and the associated decrease in effective cross-section, which enters at second order and scales approximately as $\mathrm{Ro}^{-2}$. Motivated by this scaling, we introduce an attenuation factor, $h(\lambda_p)$ that acts only on the $L_2$ nozzle term, thereby accounting for the asymmetric suppression of the anti-stellar outflow. $h(\lambda_p) = \exp(-k_{\rm tail}/\mathrm{Ro}^2)$, where $k_{\rm tail}$ is a constant calibration parameter, obtained by fitting the $\dot M_{\rm tail}$ to the mass-loss rates measured in the simulations, restricting the fit to the regime where the outflow exhibits a clear tidal-tail-like morphology. We obtain $k_{\rm tail} = 0.75$. Using this, we define the \textit{Modified Tidal Two-Tail Model} which computes the mass-loss rates as,
\begin{equation}
    \dot M_{\rm tail} = \dot M_{L_1} + h(\lambda_p) \dot M_{L_2}.
\end{equation}

This model works well in the strong-tide (low $f_\phi$), cool wind (large $\lambda_p$) regime (see blue curves in Figure \ref{fig:models_comparison}, middle). For strongly bound, cooler outflows ($\lambda_p = 7.5$), it matches the simulation results over a broad range of Roche-lobe filling ($f_\phi \lesssim 2$). At intermediate binding ($\lambda_p = 5$), good agreement is achieved only for more Roche-lobe-filling systems ($f_\phi \lesssim 0.9$). In contrast, for weakly bound, hotter winds ($\lambda_p = 2.5$), the \textit{Modified Tidal Two-Tail Model} overpredicts the mass-loss rates and can deviate by up to an order of magnitude as the planet becomes more embedded within its Roche lobe (larger $f_\phi$). The model breaks down in the weak-tide, hot-wind regime because the escape transitions to being isotropic, and the assumption that the flow is primarily channeled through the $L_1/L_2$ nozzle regions is no longer valid. 


\subsection{1D Tidal Model} \label{sec:1D_tidal model}
In the standard 1D Parker wind solution and the subsequent \textit{Modified Spherical Model} (Eqs.~\ref{eq:sonic PW} \& \ref{eq:msphere}), the tidal contribution to the effective potential is neglected when determining the sonic point. If the tidal term is retained, the sonic radius is modified and is given by \citep{1999isw..book.....L, 2009ApJ...693...23M, 2022AJ....164..234V},
\begin{equation}
r_s = a\left[ \left(\frac{M_x + (M_p/2)}{3M_\star}\right)^{\frac{1}{3}} - \left(\frac{M_x - (M_p/2)}{3M_\star}\right)^{\frac{1}{3}} \right],
\end{equation}
where
\begin{equation}
M_x = \sqrt{\left( \frac{M_p}{2}\right)^2 + \frac{8M_\star^2}{81}\left(\frac{c_s}{v_K} \right)^6},
\end{equation}
and $v_K = \sqrt{GM_\star/a}$. These tidal corrections are implemented in the latest version of the \texttt{p-winds} code, which computes the 1D wind structure using the tidally corrected sonic radius \citep{2022A&A...659A..62D}. Thus, this model is equivalent to the standard Parker wind solution with only the modification that $\Phi(r_s)$, which enters Eq.\ref{eq:PW} through $\Delta\Phi = \Phi_s - \Phi(r_s)$ for mass-loss rate calculation, is evaluated at a tidally corrected sonic radius. However, the flow is still assumed to be spherically symmetric. Hereafter, we refer to this as the \textit{1D Tidal Model}.

We compare the \textit{1D Tidal Model} with our 3D simulation results in Figure \ref{fig:models_comparison} (Right). The model broadly captures the overall trend in mass-loss rates across the full $(f_\phi, \lambda_p)$ parameter space. However, systematic deviations are present. In particular, the model tends to overpredict the mass-loss rates, with the magnitude of the discrepancy increasing toward lower $\lambda_p$ and smaller $f_\phi$, corresponding to hotter, more Roche-lobe-filling outflows. For $\lambda_p = 2.5$, it overpredicts the mass-loss rates by a factor of $\sim 1.4$-$1.6$ across the full range of $f_\phi$. At $\lambda_p = 5$, the model agrees well with the simulations at larger $f_\phi$ (less Roche-lobe filling), but the discrepancy increases toward smaller $f_\phi$, with the mass-loss rates overpredicted by up to a factor of $\sim 1.5$. For $\lambda_p = 7.5$, the model reproduces the simulation results across most of the Roche-lobe-filling range, with only modest deviations in the intermediate regime with $0.1 \lesssim f_\phi \lesssim 2$. 

While the \textit{1D Tidal Model} includes tidal corrections through a modified sonic radius, it still assumes a spherically symmetric outflow, with the effective escape area set by a sphere of radius $r_s$. However, the 3D simulations show that in the strong-tide regime the outflow becomes highly anisotropic, with escape confined to a limited solid angle around the $L_1/L_2$ points (see Figure~\ref{fig:radial profile}). As a result, the effective escape area is smaller than that implied by spherical symmetry. This mismatch in geometry leads to a corresponding overestimate of the mass flux, and likely explains the systematic overprediction of mass-loss rates by the \textit{1D Tidal Model}.
\subsection{The Mixture Model}
The results from \S\ref{sec:sph outflow formula} and \S\ref{sec:tidal tail outflow formula} suggest that the 3D simulations are naturally described by two limiting outflow geometries. The \textit{Modified Spherical Model} reproduces the simulated mass-loss rates when the escape remains approximately isotropic (large $f_\phi$, small $\lambda_p$), whereas the \textit{Modified Tidal Two-Tail Model} performs better when tides strongly channel the flow and the outflow becomes anisotropic (small $f_\phi$, large $\lambda_p$). Motivated by this behavior, and by the smooth transition between these morphologies seen in the 3D simulations, we construct a \textit{Mixture Model} that interpolates between the two limits using a single mixing weight. Since the transition is governed by changes in flow morphology that primarily alter the effective escape area and potential barrier, the resulting variation in mass-loss rate is multiplicative rather than additive. We therefore define the global mass-loss rate as a weighted geometric mean of the two models:
\begin{equation}
    \dot{M}_{\rm mix} = \dot{M}_{\rm sph}^{1-w}\dot{M}_{\rm tail}^{w} = \left(\frac{\dot{M}_{\rm tail}}{\dot{M}_{\rm sph}}\right)^w g(\lambda_p) \dot{M}_{\rm PW}.
\end{equation}
Here $w \in [0,1]$ is a scalar morphology weight that encodes the relative importance of the \textit{Modified Tidal Two-Tail Model} compared to the \textit{Modified Spherical Model}. By construction, we require $w \to 0$ in the isotropic limit and $w \to 1$ in the anisotropic tidal-tail limit. To enforce these bounds while keeping the model smooth and differentiable, we parameterize $w$
with a sigmoid (logistic) function, $w = 1/(1+ e^{-z})$, where $z$ is an unconstrained “logit” that we model as a function of the dimensionless parameters controlling the flow morphology. The simulations indicate that the transition is controlled primarily by (i) how strongly bound the atmosphere is near the planet, captured by $\lambda_p$, and (ii) how easily gas can access the $L_1/L_2$ escape channels, which we quantify with a dimensionless potential-barrier parameter, $\eta \equiv |\Phi_s - \Phi_{\rm eff}(L_1)|/c_s^2 = f_\phi \lambda_{\rm Hill}$. Small $\eta$ corresponds to a low barrier ($L_1$ funnel easily accessible), favoring a two-tail morphology, whereas large $\eta$ corresponds to a high barrier, favoring more spherical escape. Motivated by this, we take the logit to be linear in $\ln{\lambda_p}$, and $\ln({1+\eta})$:
\[
z = \beta_0 + \beta_1\ln{\lambda_p} + \beta_2\ln{(1 + \eta)} 
\]
where $\beta_0, \beta_1$, and $\beta_2$ are dimensionless calibration coefficients. We determine \{$\beta_i$\} by performing a least-squares fit in log space to the simulation results. The best-fit coefficients are, $\beta_0 = -7.00 \pm 0.69, \beta_1 = 6.13 \pm 0.59,$ and $\beta_2 = -4.03 \pm 0.54$. 

\begin{deluxetable}{lllll} \label{tab:model_summary}
\small
\setlength{\tabcolsep}{3pt}
\tablehead{
\colhead{Model} & \colhead{Key Assumptions} & \colhead{Valid Regime} & \colhead{Failure Regime} &
\colhead{Systematic Trend} 
}
\startdata
\textit{Parker Wind}
& Isotropic escape 
& \shortstack[l]{large $f_\phi$ + large $\lambda_p$}
& \shortstack[l]{small $f_\phi$; small $\lambda_p$}
& \shortstack[l]{Underpredicts at small $f_\phi$;\\overpredicts at small $\lambda_p$}\\
\\
\textit{Modified Spherical}
& Isotropic escape
& \shortstack[l]{large $f_\phi$; small $\lambda_p$}
& \shortstack[l]{small $f_\phi$; large $\lambda_p$}
& \shortstack[l]{Underpredicts when flow\\ becomes anisotropic} \\
\\
\textit{Nozzle Model}
& $L_1/L_2$ escape only 
& \shortstack[l]{small $f_\phi$; large $\lambda_p$}
& Most regimes
& \shortstack[l]{Generally overpredicts;\\best for anisotropic flows}\\
\\
\textit{Modified Tidal Tail}
& \shortstack[l]{$L_1/L_2$ escape with\\suppressed $L_2$ channel} 
& \shortstack[l]{small $f_\phi$; large $\lambda_p$}
& \shortstack[l]{large $f_\phi$; small $\lambda_p$}
& \shortstack[l]{Overpredicts when flow\\ becomes isotropic}\\
\\
\textit{1D Tidal Model}
& \shortstack[l]{Spherical escape\\geometry}
& Global trends
& \shortstack[l]{small $f_\phi$; small $\lambda_p$}
& \shortstack[l]{Modest overprediction,\\typically $\lesssim 1.6$ times}\\
\\
\textit{Mixture Model}
& Empirical interpolation
& All regimes
& --
& Matches simulations
\enddata
\caption{Summary of model performance for calculating mass-loss rates across physical regimes. “Weak/strong tide” correspond to large/small $f_\phi$, and “hot/cool winds” correspond to small/large $\lambda_p$.}
\end{deluxetable}

The \textit{Mixture Model} smoothly transitions between distinct escape regimes and captures the impact of changing outflow geometry on the mass-loss rate. When compared to 3D simulation results (Figure \ref{fig:models_comparison}, Right), it shows excellent agreement across the full parameter space and also improves upon the predictions from the \textit{1D Tidal Model}. We provide an open source implementation of this model \footnote{\url{https://github.com/sethiritika/Hydrodynamic_Atmospheric_Escape_in_Planets}} that enables a simple computation of mass-loss rates, given stellar mass, basic planetary and atmospheric properties, e.g., mass, radius, temperature \& mean molecular weight (or equivalently the sound speed). While the \textit{Mixture Model} can be extrapolated beyond the parameter space explored in this paper, its accuracy has been validated only within the range $2.5 \leq \lambda_p \leq 7.5$, probed in our simulations. The regimes of validity, systematic deviations, and key physical limitations of each model presented in this section are summarized in Table~\ref{tab:model_summary}.

\begin{figure}
    \centering
    \includegraphics[width=1.13\linewidth]{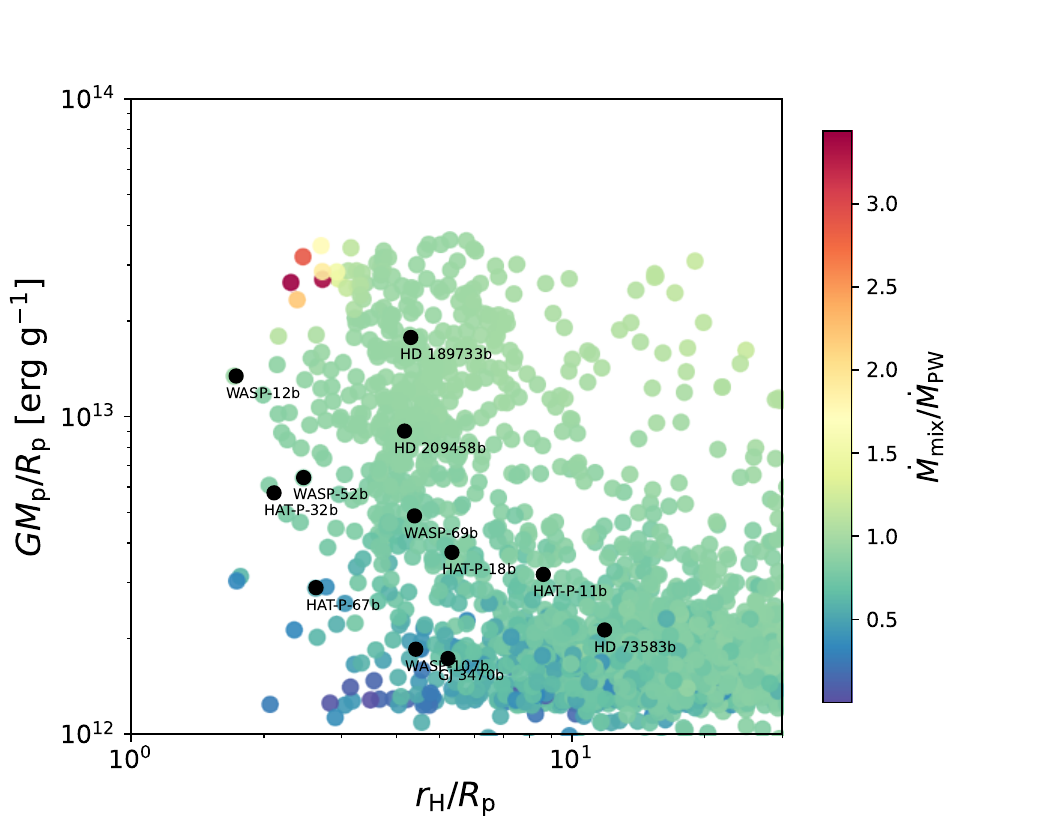}
    \caption{Ratio of Hill radius to planetary radius characterizing the degree of Roche-lobe filling vs. the planetary gravitational potential, correlated to how deeply the atmosphere is bound within the planet’s potential well. Colored points represent exoplanets with measured masses, radii, and atmospheric outflow properties from the \texttt{sunset} catalog, restricted to planets with $R_p > 1.6,R_\oplus$ \citep{2025A&A...698A.112L}. Colors indicate the predicted value of $\dot{M}_{\rm mix}/\dot{M}_{\rm PW}$. Planets with detected atmospheric escape signatures are marked with black circles.}
    \label{fig:planet_pop}
\end{figure}

\section{Applications and Comparison} \label{sec:discussion} 

\subsection{Application to the Observed Exoplanet Population}
\begin{figure}[t]
    \centering
    \includegraphics[width=\linewidth]{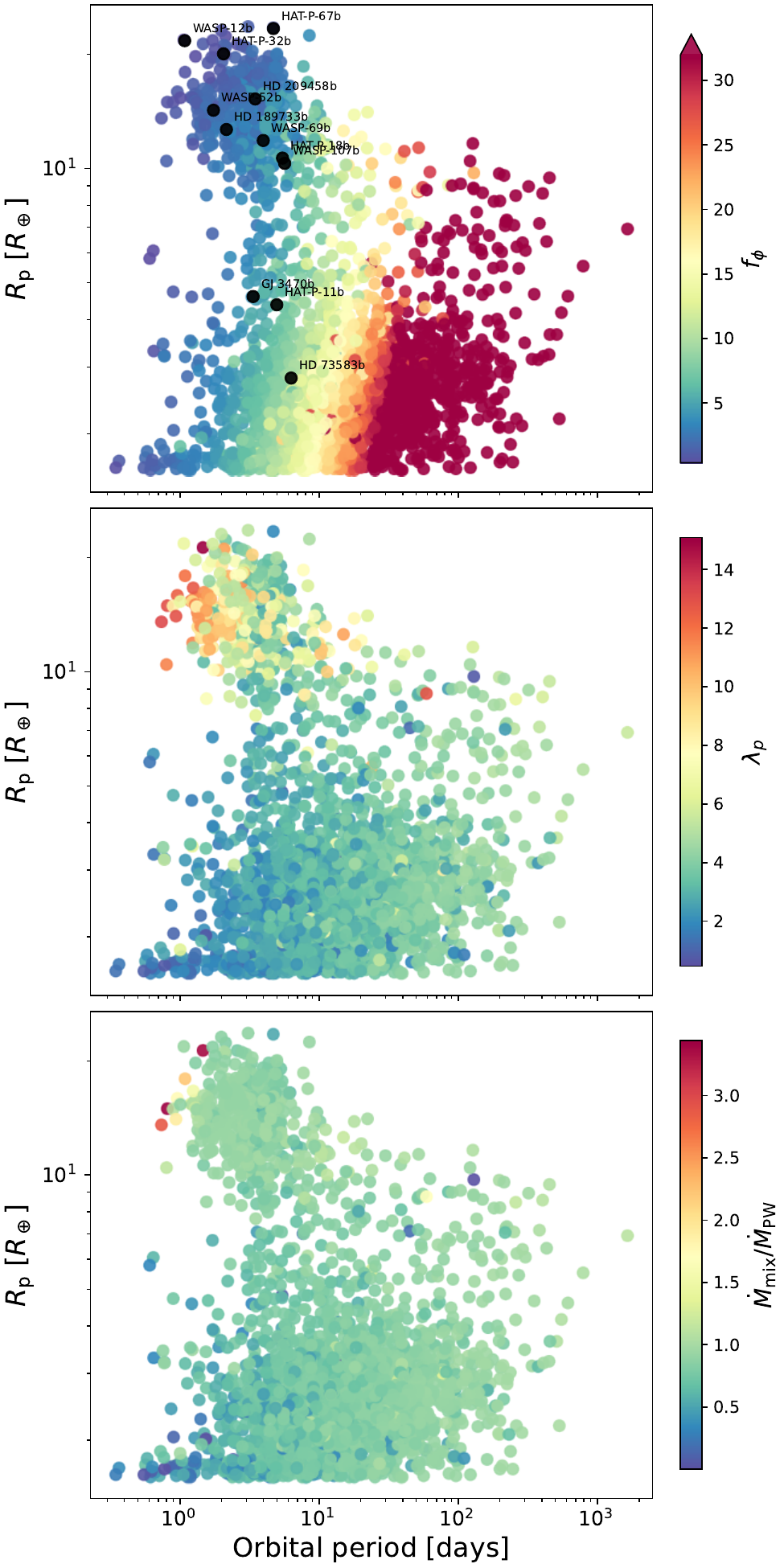}
    \caption{Observed exoplanets in $P$-$R_p$ space and color-coded by parameters connecting them to our simulations. The top panel shows $f_\phi$, which traces the degree of Roche-lobe filling; the middle panel shows the inferred $\lambda_p$, which reflects the combined effect of outflow temperature and planetary binding; and the bottom panel shows the mass-loss-rate ratio $\dot{M}_{\rm mix}/\dot{M}_{\rm PW}$, as in Figure \ref{fig:planet_pop}.}
    \label{fig:fphi_lambdap_mapping}
\end{figure}

In this section, we compare the predictions of our \textit{Mixture Model} with those of the PW model (Eq. \ref{eq:sonic PW}), which is widely used in observational studies. As discussed above, the \textit{Mixture Model} reproduces the mass-loss rates from 3D simulations with high accuracy across the full simulation grid, whereas the PW model shows systematic deviations. It underpredicts the mass-loss for cooler, more gravitationally bound outflows in the strong-tide regime, and overpredicts them for hotter, loosely bound winds. We now assess how these differences map onto the observed exoplanet population.

In Figure \ref{fig:planet_pop}, we show planets from the NASA Exoplanet Archive \citep{2013PASP..125..989A} with measured planetary mass, radius, and stellar mass. We plot the ratio of Hill radius to planetary radius, $r_H/R_p$, which characterizes the degree of Roche-lobe filling, against planetary potential, $GM_p/R_p$. For each planet, we use the \texttt{sunset} catalog of predicted photoevaporative outflow properties to estimate a characteristic outflow temperature, $T$ \citep{2025A&A...698A.112L}. The same planet population was used for an independent investigation by \citet{2025ApJ...988...63M}. Using these temperatures, we compute the sound speed as $c_s = \sqrt{\gamma k_B T/\mu m_p}$, where we set the adiabatic index, $\gamma = 5/3$, and then calculate $\lambda_p$ from Eq.~\ref{eq:lambdap} using the observed planetary masses and radii. Here, $k_B$ is the Boltzmann constant, $\mu$ is the mean molecular weight, and $m_p$ is the proton mass. 

Figure~\ref{fig:planet_pop} shows that the PW model agrees reasonably well with the \textit{Mixture Model} for most of the observed planet population, but also highlights where 1D spherical estimates should be treated with caution. We find two such observationally identifiable regimes: (i) strongly tide-affected, tightly bound systems in the small-$r_H/R_p$, large-$GM_p/R_p$ region of Figure \ref{fig:planet_pop}, where the PW model tends to underpredict the mass-loss rate; and (ii) loosely bound, hotter atmospheres in the low-$GM_p/R_p$ region, where the PW model can substantially overpredict it. Thus, $r_H/R_p$, $GM_p/R_p$, and, when possible, $\lambda_p$, provide practical diagnostics for assessing when the PW model is likely to be reliable.

To connect these quantities more directly to the demographic features of the observed exoplanet population, Figure \ref{fig:fphi_lambdap_mapping} shows the same planets in $P$-$R_p$ space, color-coded by $f_\phi$, $\lambda_p$, and the ratio $\dot{M}_{\rm mix}/\dot{M}_{\rm PW}$. This view helps identify where differences between the \textit{Mixture Model} and the PW model may matter most for interpreting planet demographics. The top panel shows the expected trend that close-in planets have smaller $f_\phi$, corresponding to stronger Roche-lobe filling, while longer-period planets have larger $f_\phi$ and approach the weak-tide, nearly isolated-planet limit. Although some long-period systems exceed the upper end of our simulated $f_\phi$ range, our simulations already reach this limiting regime, where the outflow becomes nearly isotropic and the PW model is valid. Extrapolation to still larger $f_\phi$ is therefore not expected to introduce significant additional uncertainty.

The middle panel shows the distribution of inferred $\lambda_p$ values. We find that the observed population spans a broad range, from $\lambda_p \sim 0.5$ to $15$, with roughly 81\% of planets lying within $\lambda_p \in [2.5,7.5]$, the range directly covered by our simulations. This is encouraging because it implies that the majority of the observed population lies in the regime where the \textit{Mixture Model} has been directly calibrated and tested. For planets outside this range, the model can still be evaluated, but the resulting mass-loss estimates should be interpreted as extrapolations until validated by additional simulations. The distribution of $\lambda_p$ reflects the combined effects of atmospheric heating and planetary binding energy. Close-in low-mass planets, including many super-Earths and sub-Neptunes, tend to have hot, loosely bound outflows and therefore smaller $\lambda_p$, while hot Jupiters at similar periods retain larger $\lambda_p$ because of their deeper gravitational potentials. Toward longer orbital periods, reduced irradiation lowers the characteristic outflow temperature, increasing $\lambda_p$ and producing the visible transition from hot to warm Jupiters.

The bottom panel shows the combined effect of $f_\phi$ and $\lambda_p$ on the predicted mass-loss-rate ratio.
For most observed planets, the \textit{Mixture Model} agrees closely with the PW model, with $\sim 84 \%$ of the sample having $\dot{M}_{\rm mix}/\dot{M}_{\rm PW} \in [0.7, 1.3]$. Thus, for the majority of systems, the PW model provides a reasonable first-order estimate of the mass-loss rate. However, the deviations occur in physically interpretable regions of parameter space. The \textit{Mixture Model} predicts higher mass-loss rates than the PW model for strongly tide-affected planets that are more Roche-lobe-filling (small $r_H/R_p$, small $f_\phi$) and have tightly bound atmospheres (large $GM_p/R_p$, large $\lambda_p$). This corresponds to the regime in which the PW model underpredicts the 3D simulation results. In the observed exoplanet population, however, such systems are rare, comprising $\lesssim 1\%$ of the sample and occupying the upper-left region of Figure \ref{fig:planet_pop} \& \ref{fig:fphi_lambdap_mapping}. For these planets, $\dot{M}_{\rm mix}$ exceeds $\dot{M}_{\rm PW}$ by factors of $\sim 1.5$ up to $\sim 3.5$, consistent with the expectation that strong tidal effects can produce anisotropic outflows that are not captured by the PW model.

For $\sim 15\%$ of planets, the \textit{Mixture Model} predicts lower mass-loss rates with  $\dot{M}_{\rm mix}/\dot{M}_{\rm PW} \lesssim 0.7$. These systems span a range of Roche-lobe filling or $r_H/R_p$ values but are concentrated at low planetary potential and low $\lambda_p$, corresponding to loosely bound, hot outflows. Many of these planets occupy the close-in small-planet population below the Neptune desert, where strong irradiation and shallow gravitational potentials drive hot, weakly bound winds. This is consistent with our 3D simulations, which show that the Parker wind model can overpredict mass loss in the loosely bound hot-wind regime. In the most extreme cases, $\dot{M}_{\rm mix}/\dot{M}_{\rm PW}$ reaches values as low as $10^{-5}$, implying that the Parker wind model may overestimate escape rates by several orders of magnitude. 

A few caveats are worth noting. First, the \textit{Mixture Model} is calibrated using simulations that assume a nearly isothermal outflow ($\gamma \approx 1$), whereas for the observed planet population outflows are not truly isothermal. Thus the  \texttt{sunset} temperatures used here should be interpreted as a representative value for mapping each observed system onto an effective $\lambda_p = GM_p/c_s^2R_p$, rather than as a single physical temperature. However, we verified that this mapping does not change the overall population-level trends discussed above. Second, the inferred outflow temperatures depend on assumptions about the unknown stellar spectral energy distribution, and outflow metallicity (assumed to be solar), and other modeling choices in the underlying \texttt{sunbather} framework \citep{2024A&A...688A..43L, 2025A&A...698A.112L}. If the true outflow temperatures are lower than estimated here, more planets would shift into the strongly tide-affected, tightly bound regime, increasing both the number of systems for which the PW model underpredicts mass-loss rates and the magnitude of that discrepancy. Conversely, if the true temperatures are higher, fewer systems would occupy the strong-tide regime, reducing the number of planets for which the PW model underpredicts mass loss, while increasing the fraction of systems in the loosely bound regime where it tends to overpredict. Despite these uncertainties, the qualitative trends identified here are robust.
\begin{figure*}
    \centering
    \includegraphics[width=\linewidth]{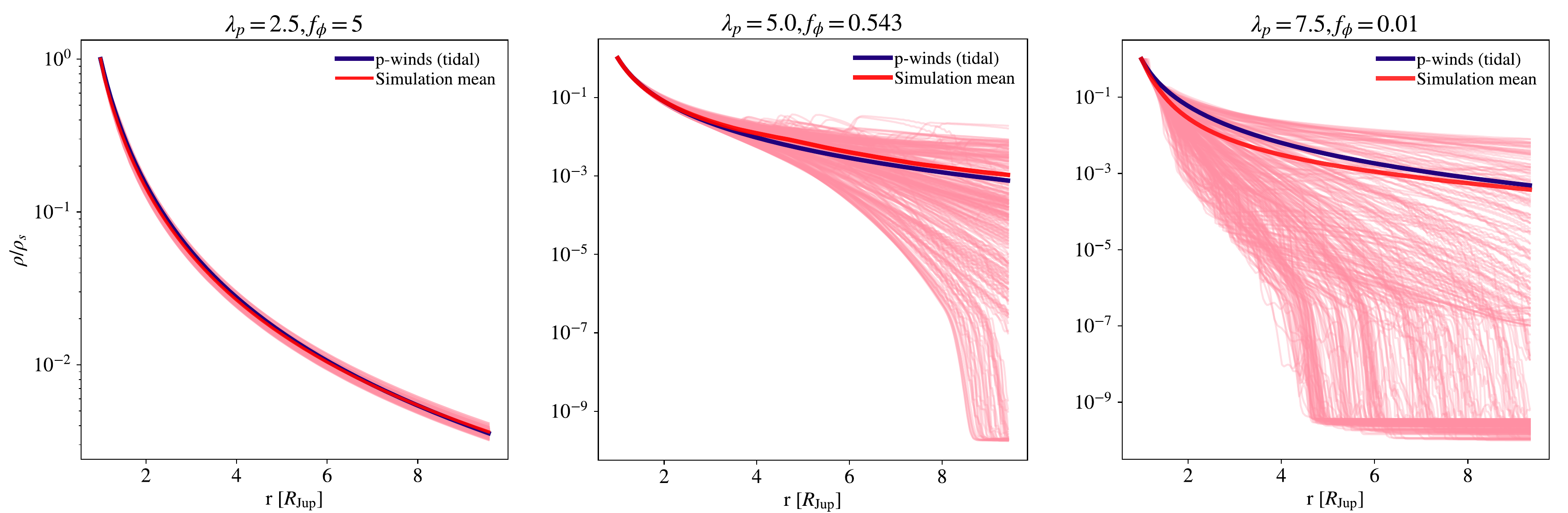}
    \caption{Radial density profiles of the outflowing atmosphere for representative simulated planets in the weak-tide, hot-wind regime (Left), an intermediate case (Middle), and the strong-tide, cooler-wind regime (Right). The densities are normalized by the planetary surface density, $\rho_s$. The blue curve shows the radial density profile predicted by the \texttt{p-winds} (tidal) code \citep{2022A&A...659A..62D, 2022ApJ...927...96V}. The pink curves show profiles extracted from our 3D simulations along rays randomly sampled from directions uniformly distributed over the sphere. The red curve denotes the mean radial density profile from simulations averaged over all sampled directions.}
    \label{fig:rho_prof}
\end{figure*}

\subsection{Radial Density Profiles from  1D Tidal Model and 3D Simulations}
The previous section focused on the PW model because of its widespread use to estimate mass-loss rates for observed systems. For comparing the structure of the outflow, however, the more relevant 1D benchmark is the \textit{1D Tidal Model}, which includes tidal corrections to the effective potential while still assuming spherical symmetry. As discussed in \S~\ref{sec:1D_tidal model}, this model provides reasonable first-order mass-loss estimates across much of our simulation grid. Here, we ask whether a 1D tidal solution that captures the integrated mass-loss rate also reproduces the radial density structure of the 3D flow.

Using the \texttt{p-winds} package, we implement the \textit{1D Tidal Model} and compute the outflow density profiles, which are inherently defined along the star–planet line (passing through the Lagrange points). We compare them to radial density profiles extracted along multiple directions from our simulations. We consider three representative cases shown in Figure \ref{fig:rho_prof}: (i) a planet deeply embedded within its Roche-lobe, with loosely-bound, hot-winds case ($f_\phi = 5$, $\lambda_p = 2.5$), where the outflow remains nearly isotropic; (ii) an intermediate case ($f_\phi = 0.543$, $\lambda_p = 5$), where the flow is nearly spherical close to the planet but becomes increasingly tidally channeled at larger radii, and (iii) a highly Roche-lobe filling and strongly bound, cooler wind case ($f_\phi = 0.01$, $\lambda_p = 7.5$), where the outflow is strongly anisotropic and exhibits a clear two-tail morphology.

We plot the density profile obtained from the \texttt{p-winds} (tidal) in Figure \ref{fig:radial profile} (blue curve), along with radial density profiles extracted along multiple directions from the simulations (light pink curves). The directional profiles are constructed by sampling rays uniformly over the sphere. We also show the mean radial density profile from the simulations (red curve), obtained by averaging the densities measured along these rays at each radial distance from the planet’s center. All density profiles are normalized by the planetary surface density, $\rho_s$.

While the mean radial density profile from the simulations nearly overlaps, the \texttt{p-winds} (tidal) profile in the weak-tides regime, the agreement decreases slightly as the flow becomes more Roche-lobe filling and more tightly bound. Nevertheless, the discrepancy remains modest, and the \textit{1D Tidal Model} continues to capture the mean radial density profile well across all regimes. However, it fails to describe the angular structure of the flow once tidal effects become significant. In the weak-tide regime, density profiles along different directions are tightly clustered, indicating that the outflow is nearly isotropic, consistent with the density maps presented in Figure \ref{fig:wind_morphology}. As a result, both the mean profile and individual rays agree closely with the 1D solution (Figure \ref{fig:radial profile}, left). In this limit, the \textit{1D Tidal Model} provides an accurate description not only of the average behavior but also of the flow along all directions.

In contrast, in the strong-tide regime, the flow becomes highly anisotropic, and density profiles vary substantially with direction. While the \textit{1D Tidal Model} is defined along the star–planet line passing through the Lagrange points, the outflow in the simulations follows curved, non-radial trajectories (see Figure \ref{fig:wind_morphology}). As a result, the density structure depends strongly on direction. Rays passing through the curved tidal tails exhibit enhanced densities relative to the \texttt{p-winds} (tidal) profile, whereas those passing through under-dense regions show a more rapid decline.

In intermediate cases, the inner flow remains approximately isotropic, while angular variations grow with radial distance from the planet’s center. The radial distance at which density profiles begin to exhibit strong directional dependence shifts inward (closer to the planet) as tidal effects strengthen (more Roche-lobe-filling configurations and cooler, more tightly bound winds). Despite these directional differences, as mentioned above, the mean density profile from the simulations remains reasonably well captured by the \textit{1D Tidal Model}, although the agreement decreases slightly in the strong-tide regime.

\section{Conclusions} \label{sec:conclusions}
In this work, we investigated the 3D structure of atmospheric escape from a Jupiter-sized planet embedded in the Roche potential of a solar-type star. To do this, we performed a suite of hydrodynamic simulations using the Athena++ framework \citep{2008ApJS..178..137S, 2020ApJS..249....4S} across a broad range of physically plausible outflow conditions, parameterized by the degree of Roche lobe filling, $f_\phi \in [0.01, 5]$, and the hydrodynamic escape parameter, $\lambda_p = 2.5, 5, 7.5$. Table \ref{tab:models} presents the full grid of 45 simulations. 
Beyond characterizing the multidimensional structure of the flow, we also compared widely used 1D mass-loss models with our 3D simulations to identify the regimes in which 1D models remain reliable and where they break down. This comparison helps clarify which aspects of atmospheric escape can be captured within a 1D framework and which require a full multidimensional treatment. Our main results are summarized as follows.

\begin{enumerate}
    \item The 3D simulations reveal a smooth morphological transition across the ($f_\phi, \lambda_p$) space. Systems in the weak-tide regime, and/or with hot, loosely bound atmospheres, corresponding to large $f_\phi$ and small $\lambda_p$, produce nearly isotropic winds. In contrast, systems in the strong-tide regime with cool, tightly bound atmospheres, corresponding to small $f_\phi$ and large $\lambda_p$ escape preferentially through the vicinity of $L_1$ and $L_2$, forming a two-tail morphology. Intermediate cases show a mixed behavior with flows remaining nearly spherical close to the planet but become increasingly tidally channeled  farther out (Figure \ref{fig:wind_morphology}).


    \item The sonic surfaces show the dynamical origin of this transition (Figure \ref{fig:mach_number}). Hot, weakly tide-affected winds pass through a nearly spherical sonic surface, consistent with a nearly isotropic escape. In cooler, more Roche-lobe-filling cases, the sonic surface is compressed along the star-planet axis and extended in transverse directions, reflecting preferential acceleration through the $L_1/L_2$	channels.
    
    
    \item The Parker wind model captures the broad mass-loss trends in the weak-tide, quasi-spherical regime, but systematically overpredicts hot, weakly bound winds and fails once tidal effects drive strongly anisotropic escape. A simulation-calibrated \textit{Modified Spherical Model} corrects this behavior in the hot-wind regime but remains limited to nearly spherical outflows and continues to fail in the strong-tide regime. 
    

    \item In the strong-tide, cool-wind regime, escape is better described as flow through the vicinity of the $L_1$ and $L_2$ regions. A \textit{Nozzle Model} captures this qualitative geometry but overpredicts the mass-loss rates, while our simulation-calibrated \textit{Modified Tidal Two-Tail Model} reproduces the 3D results in the tidally channeled regime. 
    

    \item 
    We combine these limiting descriptions into a \textit{Mixture Model}, which interpolates between quasi-spherical and tidally directed escape and reproduces the 3D mass-loss rates across the full simulation grid. Applied to observed planets, the model shows that most systems lie within the explored parameter space, supporting the broader applicability of our results. Simple diagnostics such as $r_H/R_p$, $GM_p/R_p$, and, where available, $\lambda_p$ can identify when 1D spherical estimates are reliable and when multidimensional tidal effects are important.

\end{enumerate}

More generally, our results highlight that accurately describing atmospheric escape across all tidal regimes requires more than simply modifying the effective escape energy through tidal corrections, as is typically done in the \textit{1D Tidal Model}. Tides also reshape the sonic surface and alter the effective escape area, introducing multidimensional effects that cannot be captured by spherical symmetry alone. A fully self-consistent analytic framework that accounts for sonic-surface deformation, changing escape geometry, and the coupled role of atmospheric binding and tides remains an important goal for future work. Our results therefore provide both a physically motivated framework for interpreting current observations and a foundation for improving predictive models of exoplanet atmospheric escape and evolution.

\section*{Data and Code Availability}
The mass-loss rates from all models used in this work, together with an implementation of the \textit{Mixture Model}, are publicly available on Github and archived on Zenodo: \dataset[GitHub]{https://github.com/sethiritika/Hydrodynamic_Atmospheric_Escape_in_Planets}, \dataset[doi:10.5281/zenodo.21251079]{https://doi.org/10.5281/zenodo.21251079}.\\

\section*{Acknowledgements}
We gratefully acknowledge many helpful conversations on related topics with colleagues, including A. Oklop\v{c}i\'{c}, F. Nail, R. Murray-Clay, S. Vissapragada, H. Knutson, M. Saidel and T. Hallatt.  This material is based upon work supported by the National Science Foundation under Grant No. 2306391. R.S. acknowledges the support from the Quad fellowship by International Institute of Education. This research has made use of the NASA Exoplanet Archive, which is operated by the California Institute of Technology, under contract with the National Aeronautics and Space Administration under the Exoplanet Exploration Program.

\appendix
\vspace{-1.8em}
\section{Resolution Convergence Test}  \label{app:resolution}
\begin{figure}
    \centering
    \includegraphics[width=0.6\linewidth]{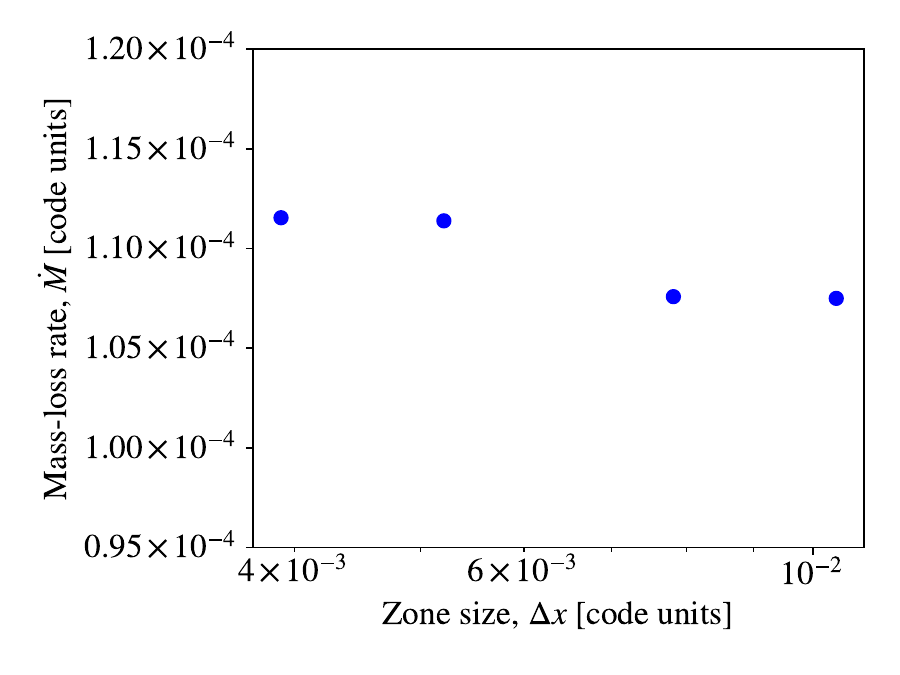}
    \caption{Resolution test for mass-loss rates from 3D hydrodynamic simulations. The points show $\dot{M}$ as a function of the smallest zone size, $\Delta x$ in the refined region for resolutions $N = 96^3$, $128^3$, $192^3$, and $256^3$. The weak dependence of $\dot{M}$ on $\Delta x$ indicates that the inferred mass-loss rate is approximately converged at resolutions coarser than what is used in the producing simulations that are analyzed in this paper.}
    \label{fig:resolution}
\end{figure}
To verify that the measured mass-loss rates from our 3D simulations are not controlled by grid resolution, we performed a resolution test varying the refined-region zone size, $\Delta x$. The tested resolutions correspond to $96
^3$, $128^3$, $192^3$, and $256^3$ zones across the refined mesh. Figure~\ref{fig:resolution} shows the resulting $\dot{M}$ as a function of 
$\Delta x$. The mass-loss rate varies only weakly over this range and is already approximately converged at resolutions coarser than the grid used for the producing the simulations analyzed in this paper. In the production simulations, the planet is embedded within one level of static mesh refinement on top of a $256^3$ base mesh, giving a $512^3$ equivalent resolution in the refined region. This corresponds to a smallest zone size, $\Delta x = 0.00195$, resolving the smallest effective planet diameter we simulated in this paper by $\sim 11$ zones.

\bibliographystyle{aasjournalv7}
\bibliography{main}

\end{document}